\documentclass[journal,draftcls,onecolumn,12pt]{IEEEtran}
\pdfoutput=1

\usepackage{amsmath}
\usepackage{amssymb}
\DeclareMathOperator*{\argmax}{arg\,max}
\DeclareMathOperator*{\argmin}{arg\,min}

\usepackage{graphicx}
\usepackage{epstopdf}
\usepackage{amsmath}
\usepackage{array}
\newcommand{\PreserveBackslash}[1]{\let\temp=\\#1\let\\=\temp}
\newcolumntype{C}[1]{>{\PreserveBackslash\centering}p{#1}}
\newcolumntype{R}[1]{>{\PreserveBackslash\raggedleft}p{#1}}
\newcolumntype{L}[1]{>{\PreserveBackslash\raggedright}p{#1}}
\usepackage[usenames]{color}
\usepackage{colortbl,booktabs}
\usepackage{tabularx}
\usepackage{multicol}
\usepackage{booktabs}
\usepackage[Symbol]{upgreek}
\usepackage{bm}
\usepackage{cite}
\usepackage{balance}
\usepackage[boxed]{algorithm2e}
\usepackage{mathrsfs}
\usepackage{url}
\usepackage{threeparttable}
\usepackage{amsmath}
\usepackage{dcolumn}
\usepackage{multirow}
\usepackage{booktabs}
\usepackage{amsfonts}

\newcommand{\tabincell}[2]{\begin{tabular}{@{}#1@{}}#2\end{tabular}}

\begin{document}

\title{Channel Feedback Based on AoD-Adaptive Subspace Codebook in FDD Massive MIMO  Systems}

\author{
	Wenqian Shen,~\IEEEmembership{Student Member,~IEEE}, Linglong Dai,~\IEEEmembership{Senior Member,~IEEE}, Byonghyo Shim,~\IEEEmembership{Senior Member,~IEEE}, Zhaocheng Wang,~\IEEEmembership{Senior Member,~IEEE}, and Robert W. Heath, Jr., ~\IEEEmembership{Fellow,~IEEE}
\thanks{A part of this paper was presented in the IEEE International Conference on Communications (ICC'17) \cite{ICC_WQShen_SubspaceCodebook}.}	
\thanks{W. Shen, L. Dai, and Z. Wang are with the Department of Electronic Engineering,
 Tsinghua University, Beijing 100084, China (E-mails: swq13@mails.tsinghua.edu.cn;
 daill@tsinghua.edu.cn; zcwang@tsinghua.edu.cn).}
\thanks{B. Shim is with the Institute of New Media and Communications, School of Electrical and Computer Engineering, Seoul National University, Seoul 151-742, Korea (e-mail: bshim@snu.ac.kr).}
\thanks{R. Heath is with the Department of Electrical and Computer Engineering, University of Texas at Austin, Austin, TX 78712-1687, USA (e-mail: rheath@utexas.edu).}
\thanks{This work was supported by the National Key Basic Research Program of China (Grant No. 2013CB329203),  the National Natural Science Foundation of China (Grant Nos. 61571270 and 61201185), the Beijing Natural Science Foundation (Grant No. 4142027), the Foundation of Shenzhen government, and the National Science Foundation (Grant No. NSF-CCF-1319556	and No. NSF-CCF-1527079).}
}
\maketitle

\begin{abstract}
Channel feedback is essential in frequency division duplexing (FDD) massive multiple-input multiple-output (MIMO) systems.
Unfortunately, previous work on multiuser MIMO has shown that the codebook size for channel feedback should scale exponentially with the number of base station (BS) antennas, which is greatly increased in massive MIMO systems.
To reduce the codebook size and feedback overhead,
we propose an angle-of-departure (AoD)-adaptive subspace codebook for channel feedback in FDD massive MIMO systems.
Our key insight is to leverage the observation that path AoDs vary more slowly than the path gains.
Within the angle coherence time, by utilizing the constant AoD information, the
proposed AoD-adaptive subspace codebook is able to quantize the channel vector in a more accurate way.
We also provide performance analysis of the proposed codebook in the large-dimensional regime,
where we prove that to limit the capacity degradation within an acceptable level, the required number of feedback bits only scales linearly with the number of resolvable (path) AoDs,
which is much smaller than the number of BS antennas.
Moreover, we compare quantized channel feedback using the proposed AoD-adaptive subspace codebook with analog channel feedback.
Extensive simulations that verify the analytical results are provided.

\end{abstract}


\IEEEpeerreviewmaketitle

\section{Introduction}\label{S1}

Massive multiple-input multiple-output (MIMO) using hundreds of base station (BS) antennas is a key technology for 5G wireless communication systems.
By simultaneously serving multiple users with simple linear precoders and combiners,
massive MIMO can improve sum spectral efficiency by orders of magnitude \cite{SPM_FRusek_ScalingupMIMO}.
Channel feedback is essential in frequency division duplex (FDD) massive MIMO to learn the channel state information at the transmitter (CSIT).
Channel feedback schemes based on the pre-defined codebook known at both BS and users have been widely used in wireless systems such as LTE/LTE-A, IEEE 802.11n/ac and WiMAX \cite{JSAC_DJLove_OverviewLimitedFeedback}.
Unfortunately, previous work on multiuser MIMO \cite{TIT_DJLove_Grassmannian},\cite{TIT_NJindal_MIMOBroadcast} has shown that the codebook size for channel feedback should scale exponentially with the number of BS antennas to guarantee the capacity loss within an acceptable level.
As the number of BS antennas in massive MIMO systems is much higher than that of current systems,
the codebook size and feedback overhead will be overwhelming.

Several channel feedback techniques have been proposed for massive MIMO systems.
Specifically, compressive sensing (CS) based channel feedback scheme exploiting the sparsity of angle-domain channel has been proposed for massive MIMO systems in \cite{WCNC_PHKuo_CSFeedback}. The channel vector is compressed into a low-dimensional measurement vector by random projection, and fed back to the BS with low overhead. Then, the BS can recover the sparse angle-domain channel via CS algorithms. 
Structured sparsity in the multi-user MIMO channel matrix can be exploited to further improve the channel recovery performance at the BS. Joint channel recovery at the BS has been proposed in \cite{TSP_XRao_DistributedCSIT} where the distributed measurement vectors of multiple users are fed back to the BS, and then the MIMO channel matrix is recovered via a joint orthogonal matching pursuit algorithm. 
Other non-CS techniques were also developed for massive MIMO channel feedback using antenna-grouping and user-grouping.
An antenna-grouping based channel feedback scheme has been proposed in \cite{TCOM_BLee_AntennaGrouping},
where multiple correlated antennas are mapped to a single representative value using predesigned patterns. Therefore, the dimension-reduced channel vector can be fed back with less overhead. 
Joint spatial division and multiplexing (JSDM) proposed in \cite{TIT_AAdhikray_JSDM} also features user-grouping.
Users with similar transmit channel covariance are grouped together and inter-group interference is mitigated through first-stage precoding based on the long-term channel statistics. 
After that, users only need to estimate and feedback the intra-group channels for second-stage precoding to mitigate intra-group interference. In this way, the overhead of both the channel training and channel feedback are reduced.
The overhead reduction of these CS-based and grouping-based channel feedback schemes is limited to the assumption of channel sparsity level and channel correlations between antennas or users.
Noncoherent trellis-coded quantization (NTCQ) has been proposed in \cite{TCOM_JChoi_Trellis} by exploiting the duality between source encoding in a Grassmannian manifold and noncoherent sequence detection. Different from traditional codebook based feedback, where the number of codewords for quantizing the channel grows exponentially with the number of antennas, the encoding complexity of NTCQ scales linearly with the number of antennas. Unfortunately, the feedback overhead with NTCQ still grows linearly with the number of antennas. 

In this paper, we propose an angle-of-departure (AoD)-adaptive subspace codebook for massive MIMO channel feedback with reduced codebook size and feedback overhead.
Our main contribution is summarized as follows:
\begin{itemize}
	\item We propose an AoD-adaptive subspace codebook with reduced codebook size and feedback overhead. We leverage the observation that path AoDs vary more slowly than path gains \cite{TVT_VVa_Beamwidth}.
During a comparably long time called ``the  angle coherence time'', which is different from the classical ``channel coherence time'', the path AoDs can be regarded as unchanged, and known to both users and BS.
Within such an angle coherence time, due to the limited scattering around the BS, the channel vector is only distributed in a subspace of the full $M$-dimensional space ($M$ is the number of BS antennas).
This subspace, which is referred to as \textit{channel subspace} in this paper, is completely determined by a limited number of path AoDs.
The quantization vectors of the proposed AoD-adaptive subspace codebook are distributed exactly on the channel subspace.
Therefore, the proposed codebook is expected to have better quantization performance.
	\item We provide performance analysis of the proposed AoD-adaptive subspace codebook in the large-dimensional regime.
Specifically, we first compute the per-user rate gap between the ideal case of perfect CSIT and the practical case of quantized channel feedback using the proposed AoD-adaptive subspace codebook.
Our result reveals that such rate gap mainly depends on the quantization error of channel vector.
Then, we derive an upper bound on the quantization error as the number of antennas grows large when the proposed AoD-adaptive subspace codebook is used.
Finally, we show that the required number of feedback bits to ensure a constant rate gap only scales linearly with the number of resolvable paths,
which is much smaller than the number of BS antennas.
Moreover, we compare quantized channel feedback using the proposed AoD-adaptive subspace codebook with analog channel feedback.
All the analytical results are verified by extensive simulations\footnote{Simulation codes are provided to reproduce the results presented in this paper: \url{http://oa.ee.tsinghua.edu.cn/dailinglong/publications/publications.html}.}.
\end{itemize}

The most related work to this paper is the channel statistics-based codebooks \cite{TVT_DJLove_LimitedFeedback},\cite{JCN_MSSim_CScodebook}.
A rotated codebook based on the channel statistics was proposed in \cite{TVT_DJLove_LimitedFeedback} to quantize the spatially correlated channel vectors.
In that approach, the required number of feedback bits scales linearly with the rank of transmit channel correlation matrix \cite{GLOBECOM_BClerckx_StaticticsCodebook}.
Another feedback technique was proposed in \cite{JCN_MSSim_CScodebook} under the framework of compressive sensing, where a codebook was designed to quantize the low-dimensional measurement vector, which will be fed back to the BS and then utilized to recover the channel vector. By jointly considering the channel statistics and the sensing matrix, the proposed codebook in \cite{JCN_MSSim_CScodebook} can quantize the measurement vector and then feed them back with reduced overhead. Both the conventional channel statistics-based codebook and our proposed AoD-adaptive subspace codebook have modified the distribution of quantization vectors based on some prior information such as the channel statistics, the sensing matrix, and the AoDs to achieve better quantization performance. 
But work is different from these channel statistics-based codebooks, as we exploit the slow-varying AoD information and design an AoD-adaptive subspace codebook in the angle coherence time. The quantization vectors of the proposed AoD-adaptive subspace codebook are exactly distributed on the channel subspace in the angle coherence time, and thus can quantize the channel vectors with better performance. Our paper differs from our previous work \cite{ICC_WQShen_SubspaceCodebook} in codebook design and performance analysis. We expand the proposed AoD-adaptive subspace codebook to a more general scenario with uniform planar array (UPA). We also propose an AoD acquisition method to further improve our codebook design scheme. Finally,  we add performance comparison between quantized channel feedback using the proposed AoD-adaptive subspace codebook with analog channel feedback to show the superior performance of the proposed codebook.

 The rest of the paper is organized as follows.
 Section~\ref{S2} presents the system model including the massive MIMO downlink channel model, channel feedback procedure, and downlink precoding.
 In Section~\ref{S3}, we review the angle coherence time and present the proposed AoD-adaptive subspace codebook. Finally, we describe the AoD acquisition method.
 Performance analysis of the proposed AoD-adaptive subspace codebook in terms of sum rate is provided in Section~\ref{S4}. We compare quantized channel feedback using the proposed codebook with analog channel feedback in Section~\ref{S5}.
 Section~\ref{S6} shows the simulation results to verify the analytical results.
 Our conclusions are drawn in Section~\ref{S6}.

 \textit{Notation}: Boldface capital and lower-case letters stand for
 matrices and vectors, respectively.
 The transpose, conjugate, conjugate transpose, and inverse of a matrix are denoted by $(\cdot)^\text{T}$, $(\cdot)^*$, $(\cdot)^\text{H}$ and $(\cdot)^{-1}$, respectively.
 $\mathbf{H}^{\dagger}=\mathbf{H}(\mathbf{H}^{\text{H}}\mathbf{H})^{-1}$ is the Moore-Penrose pseudo-inverse of $\mathbf{H}$.
 $\otimes$ is the Kronecker product operator.
 $\|\mathbf{h}\|$ and $|s|$ are the \text{norm} of a vector and the absolute value of a scalar.
 $\measuredangle(\mathbf{x},\mathbf{y})$ is the angle between $\mathbf{x}$ and $\mathbf{y}$, and $\sin^2(\measuredangle(\mathbf{x},\mathbf{y}))=1-\frac{|\mathbf{x}^\text{H}\mathbf{y}|^2}{\|\mathbf{x}\|^2\|\mathbf{y}\|^2}$.
 $\text{E}\left[\cdot\right] $ denotes the expectation operator.
 Finally, $\mathbf{I}_P$ denotes the identity matrix of size $P\times P$.

\section{Massive MIMO System Model}\label{S2}
In this section, we introduce the massive MIMO downlink channel model
and the channel feedback procedure.
To quantify the performance of our strategy, we review the per-user rate calculated assuming zero-forcing (ZF) precoding based on the fed back CSI.
\subsection{Massive MIMO Downlink Channel Model}\label{S2.1}
\begin{figure}[t!]
\vspace*{-2mm}
\begin{center}
 \includegraphics[width=0.7\textwidth, keepaspectratio]{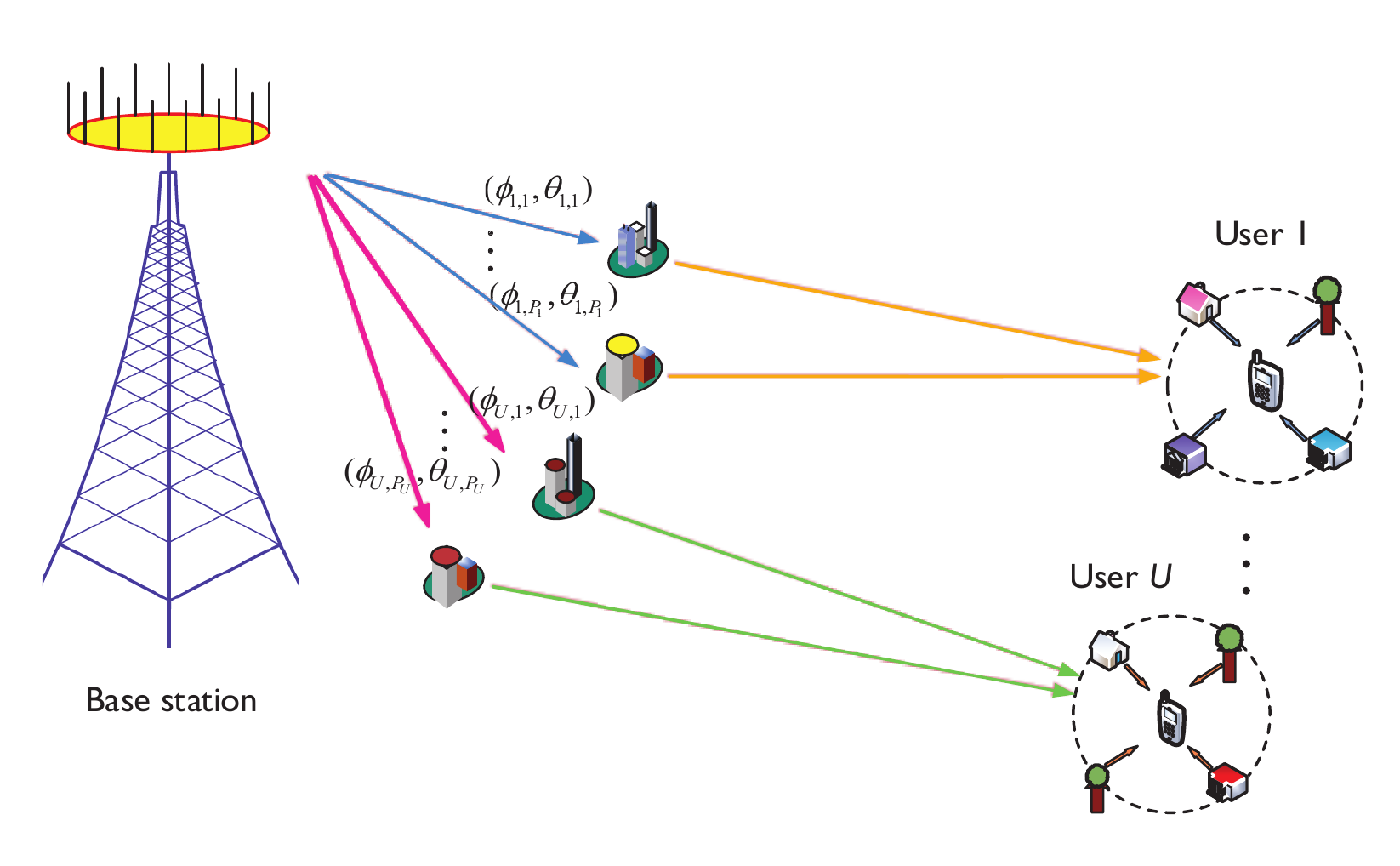}
\end{center}
\vspace*{-4mm}
\caption{Ray-based channel model. The channel between the BS and user $u$ is composed of $P_u$ resolvable propagation paths, each of which is characterized by path gain $g_{u,i}$ and path AoD $(\phi_{u,i},\theta_{u,i})$ with $i=1,2,\cdots,P_u$.}
\label{Fig1} 
\end{figure}

 In this paper, we consider a massive MIMO system with $M$ antennas at the BS and $U$ users ($M\gg U$) \cite{SPM_FRusek_ScalingupMIMO},
 where the BS is equipped with a uniform linear array (ULA) or uniform planar array (UPA) of antennas, and a single antenna is used by each user.
 We adopt the classical narrowband ray-based channel model \cite{CambridgeUP_TDavid_Fundamentals},\cite{JSSP_AAhmed_EstimationHybrid} as shown in Fig. \ref{Fig1}.
 The downlink channel vector $\mathbf{h}_u\in\mathbb{C}^{M\times 1}$ for the $u$-th user can be described as
 \begin{align}\label{eqhk1} 
 \mathbf{h}_u=\sum_{i=1}^{P_u}g_{u,i}\mathbf{a}(\phi_{u,i},\theta_{u,i}),
 \end{align}
 where $P_u$ is the number of resolvable paths from the BS to the $u$-th user,
 $g_{u,i}$ is the complex gain of the $i$-th propagation path of the $u$-th user, which is identically and independently distributed (IID) with zero mean and unit variance,
 and $\phi_{u,i}$ and $\theta_{u,i}$ are the azimuth and elevation AoDs of the $i$-th propagation path of the $u$-th user, respectively.
 The steering vector $\mathbf{a}(\phi_{u,i},\theta_{u,i})$ denotes the antenna array response of the $i$-th path of the $u$-th user. For a ULA, the array response
 \begin{align}\label{eq_ste_vec}
 \mathbf{a}(\phi_{u,i})=\frac{1}{\sqrt{M}}\left[1,e^{j2\pi\frac{d}{\lambda}\sin\phi_{u,i}},\cdots, e^{j2\pi\frac{d}{\lambda}(M-1)\sin\phi_{u,i}}\right]^\text{T},
 \end{align}
 where $d$ is the antenna spacing at the BS, $\lambda$ is the wavelength of the carrier frequency.
 For a UPA with $M_1$ horizontal antennas and $M_2$ vertical antennas ($M=M_1\times M_2$), the array response
  \begin{align}\label{eq_ste_vec_UPA}
 \mathbf{a}(\phi_{u,i},\theta_{u,i})=&\frac{1}{\sqrt{M}}\left[1,e^{j2\pi\frac{d}{\lambda}\cos\theta_{u,i}\sin\phi_{u,i}},\cdots, e^{j2\pi\frac{d}{\lambda}(M_1-1)\cos\theta_{u,i}\sin\phi_{u,i}}\right]^\text{T}\otimes \\\nonumber
 &\left[1,e^{j2\pi\frac{d}{\lambda}\sin\theta_{u,i}},\cdots, e^{j2\pi\frac{d}{\lambda}(M_2-1)\sin\theta_{u,i}}\right]^\text{T}.
 \end{align}
 In matrix form with $\mathbf{A}_u=[\mathbf{a}(\phi_{u,1},\theta_{u,1}),\mathbf{a}(\phi_{u,2},\theta_{u,2}),\cdots,\mathbf{a}(\phi_{u,P_u},\theta_{u,P_u})] \in\mathbb{C}^{M \times P_u}$ and $\mathbf{g}_u=[g_{u,1},g_{u,2},\cdots,g_{u,P_u}]^\text{T}\in\mathbb{C}^{P_u\times 1}$, we have
  \begin{align}\label{eqhk2}
 \mathbf{h}_u=\mathbf{A}_u\mathbf{g}_u.
 \end{align}
 Further, we can denote the concatenation of channel vectors for all $U$ users as $\mathbf{H}=[\mathbf{h}_1,\mathbf{h}_2,\cdots,\mathbf{h}_U]
 \in\mathbb{C}^{M\times U}$.

\subsection{Quantized Channel Feedback}\label{S2.2}
The downlink channel information in (\ref{eqhk1}) can be obtained at the user side through downlink channel training.
Although the training overhead to obtain the downlink channel vector is increased in massive MIMO systems, there are effective downlink training methods with reduced training overhead \cite{JSSP_AAhmed_EstimationHybrid,TVT_WShen_JointChannelFeedback}. Thus, in this paper, each user is assumed to know its own channel vector.

Channel vector $\mathbf{h}_u$ in (\ref{eqhk1}) is also required by the BS to perform power allocation and precoding,
which is usually realized by channel feedback from users to the BS.
The channel vector $\mathbf{h}_u$ is first quantized and then fed back to the BS.
The quantization of $\mathbf{h}_u\in\mathbb{C}^{M\times 1}$ at the $u$-th user is performed by the quantization codebook $\mathcal{C}_u=\{\mathbf{c}_{u,1},\mathbf{c}_{u,2},\cdots,\mathbf{c}_{u,2^B}\}$,
which consists of $2^B$ different $M$-dimensional unit-norm column vectors, where $B$ is the number of feedback bits.
The detailed codebook design will be discussed later in Section \ref{S3}.
The quantization codebook $\mathcal{C}_u$ is known to both the BS and the $u$-th user.
The $u$-th user quantizes its channel vector $\mathbf{h}_u$ to a quantization vector $\mathbf{c}_{u,i_u}$, where the quantization index $i_u$ is computed according to
\begin{align} \label{eqFu}
i_u=\argmin_{i\in\{1,2,\cdots,2^B\}}\sin^2(\measuredangle(\mathbf{h}_u,\mathbf{c}_{u,i}))=\argmax_{i\in\{1,2,\cdots,2^B\}}|\mathbf{\tilde{h}}_u^\text{H}\mathbf{c}_{u,i}|^2,
\end{align}
where $\mathbf{\tilde{h}}_u=\frac{\mathbf{h}_u}{\|\mathbf{h}_u\|}$ is the channel direction.

Note that only the channel direction $\mathbf{\tilde{h}}_u$ is quantized,
while the channel magnitude $\|\mathbf{h}_u\|$ is not quantized by the codebook $\mathcal{C}_u$.
Channel magnitude information can be used to allocate power and rate across multiple channels,
but it is just a scalar value and thus is also easy to fed back.
We follow the common assumption that the channel magnitude can be fed back to the BS perfectly so that we focus on the quantization of channel direction,
which is more challenging for channel feedback.

The index $i_u$ can be fed back from the $u$-th user to the BS through $B$ dedicated bits.
After receiving these $B$ bits (thus the index $i_u$), the BS can generate the channel vector through quantized channel feedback $\hat{\mathbf{h}}_u=\|\mathbf{h}_u\|\mathbf{c}_{u,i_u}$.
The concatenation of the fed back channel vectors can be denoted as $\mathbf{\hat{H}}=[\mathbf{\hat{h}}_1,\mathbf{\hat{h}}_2,\cdots,\mathbf{\hat{h}}_U] \in\mathbb{C}^{M\times U}$.

\subsection{Per-User Rate}\label{S2.3}
The BS can perform downlink precoding to eliminate interference among multiple users based on the channel matrix $\mathbf{\hat{H}}$ through quantized channel feedback.
In this paper, we consider ZF precoding.
The transmit signal $\mathbf{x}\in\mathbb{C}^{M\times 1}$ after ZF precoding is given by
\begin{align}
\mathbf{x}=\sqrt{\frac{\gamma}{U}}\mathbf{\hat{V}}\mathbf{s},
\end{align}
where $\gamma$ is the transmit power,
$\mathbf{s}=[s_1,s_2,\cdots,s_U]^\text{T}\in\mathbb{C}^{U\times 1}$ is the signals intended for $U$ users with the normalized power $\text{E}\left[|s_i|^2\right] =1$,
and $\mathbf{\hat{V}}=[\mathbf{\hat{v}}_1,\mathbf{\hat{v}}_2,\cdots,\mathbf{\hat{v}}_U]\in\mathbb{C}^{M\times U}$ is the ZF precoding matrix consisting of $U$ different $M$-dimensional unit-norm precoding vectors $\mathbf{\hat{v}}_i\in\mathbb{C}^{M\times 1}$, which is obtained as the normalized $i$-th column of ${\mathbf{\hat{H}}}^\dagger$, i.e., $\mathbf{\hat{v}}_i=\frac{{\mathbf{\hat{H}}}^\dagger(:,i)}{\|{\mathbf{\hat{H}}}^\dagger(:,i)\|}$.

 After the channel, the received signal at the $u$-th user can be described as
 \begin{align}
y_u&=\mathbf{h}_u^\text{H}\mathbf{x}+n_u\\\nonumber
&=\sqrt{\frac{\gamma}{U}}\mathbf{h}_u^\text{H}\mathbf{\hat{v}}_us_u+\sqrt{\frac{\gamma}{U}}\sum_{i=1,i\neq u}^{U}\mathbf{h}_u^\text{H}\mathbf{\hat{v}}_is_i+n_u,
\end{align}
where $n_u$ is the complex Gaussian noise at the $u$-th user with zero mean and unit variance.
Thus, the signal-to-interference-plus-noise ratio (SINR) at the $u$-th user is
 \begin{align}
\text{SINR}_u=\frac{\frac{\gamma}{U}|\mathbf{h}_u^\text{H}\mathbf{\hat{v}}_u|^2}{1+\frac{\gamma}{U}\sum_{i=1,i\neq u}^{U}|\mathbf{h}_u^\text{H}\mathbf{\hat{v}}_i|^2}.
\end{align}
Assuming Gaussian signaling and knowledge of SINR, the per-user rate $R_\text{Quantized}$ with quantized channel feedback is
 \begin{align}\label{eq_R_quantized}
R_\text{Quantized}&=\text{E}\left[ \log_2(1+\text{SINR}_u)\right]  \\\nonumber
&=\text{E}\left[\log_2\left(1+\frac{\frac{\gamma}{U}|\mathbf{h}_u^\text{H}\mathbf{\hat{v}}_u|^2}{1+\frac{\gamma}{U}\sum_{i=1,i\neq u}^{U}|\mathbf{h}_u^\text{H}\mathbf{\hat{v}}_i|^2}\right)\right].
\end{align}

The per-user rate $R_\text{Quantized}$ depends on the precoding matrix $\mathbf{\hat{V}}$,
which is affected by the channel matrix $\mathbf{\hat{H}}$ through quantized channel feedback.
In the following Section \ref{S3},
we present the proposed AoD-adaptive subspace codebook to provide the reliable channel feedback with low overhead.

\section{Proposed AoD-Adaptive Subspace Codebook}\label{S3}
In this section, we review the angle coherence time and propose an AoD-adaptive subspace codebook.
Then, we explain how to obtain the AoDs for the BS and users during the angle coherence time.

\subsection{AoD-Adaptive Subspace Codebook}\label{S3.2}
We observe from Fig. \ref{Fig1} that the $i$-th resolvable path of the $u$-th user with path AoD $(\phi_{u,i},\theta_{u,i})$ and path gain $g_{u,i}$ is composed of a number of unresolvable paths, each of which is generated by a scatter surrounding the $u$-th user.
The path AoD $(\phi_{u,i},\theta_{u,i})$ in (\ref{eqhk1}) mainly depends on the surrounding obstacles around the BS, which may not physically change their position in much longer time than the channel coherence time.
On the contrary, the path gain $g_{u,i}$ seen by the BS and the $u$-th user depends on a number of unresolvable paths, each of which are generated by scatters around the $u$-th user.
Therefore, the path gains vary much faster than path AoDs \cite{TVT_VVa_Beamwidth}.
Accordingly, the angle coherence time, during which the path AoDs can be regarded as static, is much longer than the channel coherence time.

During the angle coherence time, the channel vector $\mathbf{h}_u$ is distributed in the \textit{channel subspace}.
As shown in (\ref{eqhk1}) and (\ref{eqhk2}), the channel vector $\mathbf{h}_u$ between the BS and the $u$-th user is composed of $P_u$ paths as $\mathbf{h}_u=\sum_{i=1}^{P_u}g_{u,i}\mathbf{a}(\phi_{u,i},\theta_{u,i})=\mathbf{A}_u\mathbf{g}_u$.
Therefore, the channel vector $\mathbf{h}_u$ is actually distributed on the column space of $\mathbf{A}_u\in\mathbb{C}^{M \times P_u}$,
which is formed by linear combination of $\mathbf{A}_u$'s column vectors.
For example, due to the limited scattering of millimeter-wave (mmWave) signals, the number of paths $P_u$ is much smaller (e.g., $2\sim 8$ for 6-60GHz \cite{Access_TSRappaport_mmwave}) than the number of BS antennas $M$ (e.g., $M=128,256$).
We also expect this to be true even with massive MIMO since the number of resolvable paths $P_u$ seen by the BS depends on the clusters of scatters around the BS whose number is usually limited.
Therefore, the column space of $\mathbf{A}_u$, which is completely dependent on path AoDs $\{(\phi_{u,i},\theta_{u,i})\}_{i=1}^{P_u}$, is only a subspace of the full $M$-dimensional space.
This subspace is referred to as channel subspace in this paper.

\begin{figure}[t!]
\vspace*{-2mm}
\begin{center}
 \includegraphics[width=0.8\columnwidth, keepaspectratio]{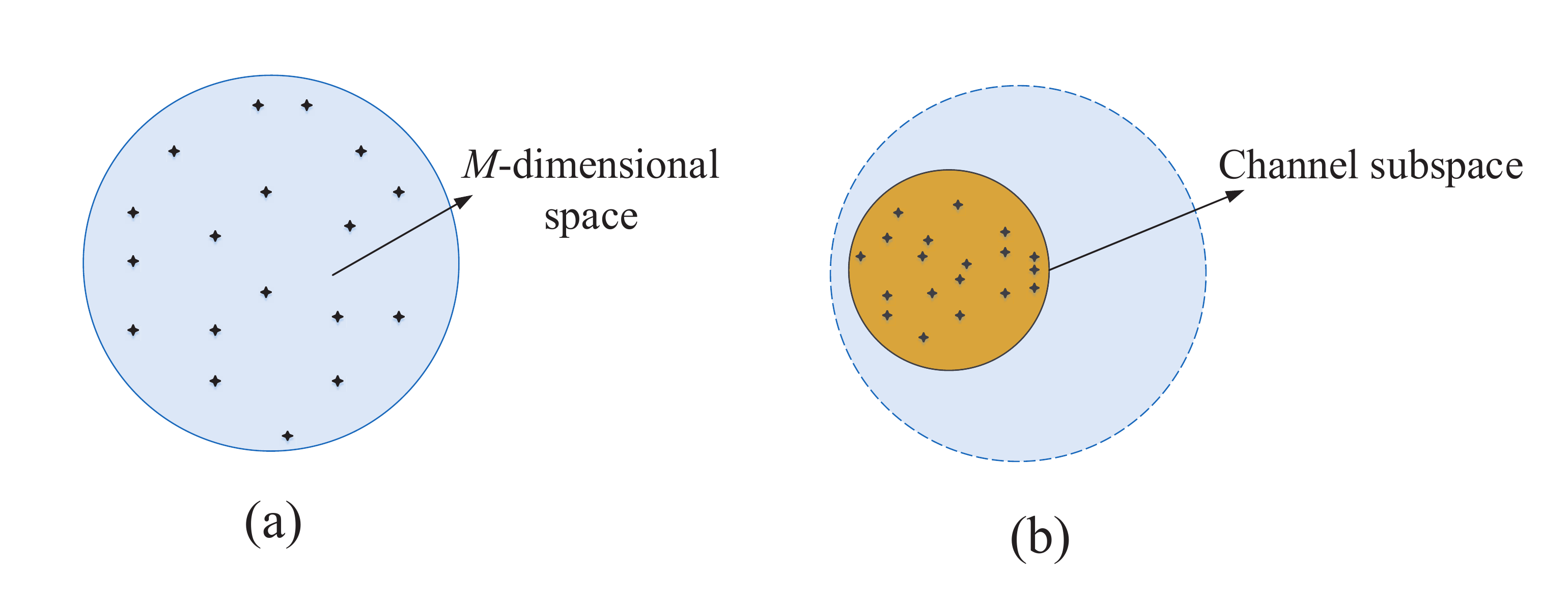}
\end{center}
\vspace*{-4mm}
\caption{Codebook comparison: (a) the classical RVQ-based codebook; (b) the proposed AoD-adaptive subspace codebook.}
\label{Fig2} 
\end{figure}

In this paper, we propose an AoD-adaptive subspace codebook where the quantization vectors are exactly distributed on the channel subspace as shown in Fig. \ref{Fig2}.
In this subsection, we assume that the quantized AoDs can be obtained at both the BS and the $u$-th user denoted by $\{(\hat{\phi}_{u,1},\hat{\theta}_{u,1}),(\hat{\phi}_{u,2},\hat{\theta}_{u,2}),\cdots,(\hat{\phi}_{u,P_u},\hat{\theta}_{u,P_u})\}$;
how to obtain the AoDs will be discussed later in next Subsection \ref{S3.3}.
Once the BS and the $u$-th user obtain the quantized AoDs,
they can generate the steering matrix $\hat{\mathbf{A}}_u=[\mathbf{a}(\hat{\phi}_{u,1},\hat{\theta}_{u,1}),\mathbf{a}(\hat{\phi}_{u,2},\hat{\theta}_{u,2}),\cdots,\mathbf{a}(\hat{\phi}_{u,P_u},\hat{\theta}_{u,P_u})]\in\mathbb{C}^{M\times P_u}$.
Then, the quantization vector $\mathbf{c}_{u,i}$ of the proposed AoD-adaptive subspace codebook $\mathcal{C}_{k}=\{\mathbf{c}_{u,1},\mathbf{c}_{u,2},\cdots,\mathbf{c}_{u,2^{B}}\}$ is generated as
 \begin{align}\label{eq_cki}
\mathbf{c}_{u,i}=\hat{\mathbf{A}}_u\mathbf{w}_i,
\end{align}
where unit-norm vector $\mathbf{w}_i\in\mathbb{C}^{P_u\times 1}$ is assumed to be a codeword in an analytical codebook, i.e., random vector quantization (RVQ)-based codebook, which is randomly generated by selecting vectors independently from the uniform distribution on the complex $M$-dimensional unit sphere \cite{TIT_NJindal_MIMOBroadcast,WCNC_DJRyan_PerformanceRVQ}. 
We consider $\mathbf{w}_i$ under the RVQ framework is to enable the performance analysis of the proposed AoD-adaptive subspace codebook. 
In practice, $\mathbf{w}_i$ can be chosen from the optimal vector quantization codebook generated by Lloyd algorithm \cite{TCOM_YLinde_LLoydLBG}.
Moreover, the performance of the proposed AoD-adaptive subspace codebook under RVQ framework is very close to that under Lloyd-based framework, which will be shown through simulations.

Note that the quantization vector $\mathbf{c}_{u,i}$ is a unit-norm vector $\|\mathbf{c}_{u,i}\|=\|\hat{\mathbf{A}}_u\mathbf{w}_i\|=1$ which can be proved as
\begin{align} \label{eq_ci_norm}
\|\hat{\mathbf{A}}_u\mathbf{w}_i\|^2&=\|\Sigma_{p=1}^{P_u}\mathbf{a}(\hat{\phi}_{u,p},\hat{\theta}_{u,p})w_{i,p}\|^2 \\\nonumber
&\overset{M \rightarrow \infty}{=}\Sigma_{p=1}^{P_u}\|\mathbf{a}(\hat{\phi}_{u,p},\hat{\theta}_{u,p})w_{i,p}\|^2 \\\nonumber
&=\Sigma_{p=1}^{P_u}|w_{i,p}|^2 \\\nonumber
&=\|\mathbf{w}_i\|^2=1,
\end{align}
where the second equation is true due to the orthogonality among column vectors $\mathbf{a}(\hat{\phi}_{u,p},\hat{\theta}_{u,p})$ of $\hat{\mathbf{A}}_u$ when $M\rightarrow\infty$ assuming that the AoDs are distinguished enough (see Appendix I).

The quantization vector $\mathbf{c}_{u,i}=\hat{\mathbf{A}}_u\mathbf{w}_i$ in the proposed AoD-adaptive subspace codebook $\mathcal{C}_u$ is distributed in the column space of $\hat{\mathbf{A}}_u$,
which only depends on the quantized AoDs $\{(\hat{\phi}_{u,1},\hat{\theta}_{u,1}),(\hat{\phi}_{u,2},\hat{\theta}_{u,2}),\cdots,(\hat{\phi}_{u,P_u},\hat{\theta}_{u,P_u})\}$.
If the BS and the $u$-th user obtain the exact AoDs,
i.e., $\hat{\mathbf{A}}_u=\mathbf{A}_u$ (which can be proved to be possible at expense of a small amount of additional overhead in next subsection \ref{S3.3}),
the quantization vectors $\mathbf{c}_{u,i}$ will be distributed exactly on the channel subspace.
Note that the proposed codebook is AoD adaptive, i.e., the quantization vector $\mathbf{c}_{u,i}$ in (\ref{eq_cki}) can be updated according to the AoDs in current angel coherence time.
Since the angle coherence time is much longer than channel coherence time, the update frequency of the proposed AoD-adaptive subspace codebook is low.


\subsection{AoD Acquisition}\label{S3.3}
In this subsection, we will describe how the BS and users obtain AoDs within the angle coherence time.
As presented in Subsection \ref{S2.2}, the channel vectors are assumed to have been estimated at the users.
Users estimate AoDs with high angle resolution      from channel vectors by using the multiple signal classification (MUSIC) algorithm \cite{TAP_RSchmidt_MUSIC}.
Specifically, the correlation matrix of channel vector $\mathbf{h}_u$ of the $u$-th user can be calculated as $\mathbf{R}_u=\text{E}[\mathbf{h}_u\mathbf{h}_u^\text{H}]$ where the expectation is estimated using the sample average obtained over several channel estimates at the $u$-th user during the angle coherence time. $\mathbf{R}_u$ has $P_u$ dominant eigenvectors associated with $P_u$ propagation paths and $M-P_u$ eigenvectors associated with noise. Then, we can construct the $M\times (M-P_u)$ noise space $\mathbf{N}_u$ spanned by the noise eigenvectors. Since the noise eigenvectors are orthogonal to the steering vectors of $P_u$ propagation paths, the path AoDs can be estimated at the peak points of the spatial spectrum function of the noise space $\mathbf{N}_u$ which is given by 
\begin{align} \label{eq_Pphi}
\mathcal{P}(\phi,\theta) =\frac{1}{\mathbf{a}(\phi,\theta)^\text{H}\mathbf{\textsc{N}}_u\mathbf{N}_u^\text{H}\mathbf{a}(\phi,\theta)}.
\end{align}
Since the BS also needs to know AoDs to generate the AoD-adaptive subspace codebook $\mathcal{C}_u$ in (\ref{eq_cki}),
the estimated AoDs have to be fed back to the BS. A common uniform quantization can be adopted to quantize the AoDs with $B_0$ bits for each AoD. Since the angle coherence time is much longer than the channel coherence time, the average overhead for AoD feedback is low. Due to limitations, we leave other advanced AoD quantization and feedback schemes for future work.
Next, we will focus on the quantitative performance analysis of the proposed AoD-adaptive subspace codebook.

%

\section{Performance Analysis of The Proposed Subspace Codebook}\label{S4}
In this section, we calculate the rate gap between the ideal case of perfect CSIT and the practical case of quantized channel feedback using a RVQ framework in the large-dimensional regime, i.e, the number of the BS antennas $M\rightarrow \infty$.
We then analyze the quantization error of the proposed structure which dominates the rate gap of quantized channel feedback.
Finally, we derive a lower bound of the required number of feedback bits to limit the rate gap within a constant value.
\subsection{Rate Gap}\label{S4.1}
In the ideal case of perfect CSIT at the BS, i.e., $\mathbf{\hat{H}}=\mathbf{H}$,
the ZF precoding vector $\mathbf{v}_{\text{Ideal},i}\in\mathbb{C}^{M\times 1}$ is obtained as the normalized $i$-th column of $\mathbf{H}^\dagger$.
Thus, $\mathbf{v}_{\text{Ideal},i}$ is orthogonal to the $u$-th user's channel vector $\mathbf{h}_u$ for any $i\neq u$,
i.e., the inter-user interference $|\mathbf{h}_u^\text{H}\mathbf{v}_{\text{Ideal},i}|=0$.
The corresponding per-user rate is
 \begin{align} \label{eq_R_ideal}
R_\text{Ideal}=\text{E}\left[\log_2\left(1+\frac{\gamma}{U}|\mathbf{h}_u^\text{H}\mathbf{v}_{\text{Ideal},k}|^2\right)\right].
\end{align}
In the practical case of quantized channel feedback, the BS can only obtain the fed back channel $\mathbf{\hat{H}}$,
which is not identical with the ideal channel $\mathbf{H}$.
The ZF precoding is performed based on $\mathbf{\hat{H}}$,
and the precoding vector $\mathbf{\hat{v}}_i \in \mathbb{C}^{M\times 1}$ is obtained as the normalized $i$-th column of $\mathbf{\hat{H}}^\dagger$.
Since the inter-user interference is nonzero now (i.e., $|\mathbf{h}_u^\text{H}\mathbf{\hat{v}}_i|\neq 0$), the degraded per-user rate $R_\text{Quantized}$ is shown in (\ref{eq_R_quantized}).
Following Theorem 1 of \cite{TIT_NJindal_MIMOBroadcast}, the rate gap $\Delta R_\text{Quantized} = R_\text{Ideal}-R_\text{Quantized}$ can be upper bounded as
\begin{align}\label{delatR1}
\Delta R_\text{Quantized}\leq \log_2\left(1+(U-1)\frac{\gamma}{U}\text{E}\left[\|\mathbf{h}_u\|^2\right] \text{E}\left[|\tilde{\mathbf{h}}_u^\text{H}\mathbf{\hat{v}}_i|^2\right] \right),
\end{align}
where $\tilde{\mathbf{h}}_u$ is the normalized channel vector in (\ref{eqFu}).
The average multi-user interference $\text{E}\left[|\tilde{\mathbf{h}}_u^\text{H}\mathbf{\hat{v}}_i|^2\right] $ in (\ref{delatR1}) can be upper bounded in the following \textbf{Lemma 1} in the large-dimensional regime.

\textbf{Lemma 1}: The multi-user interference $\text{E}\left[|\tilde{\mathbf{h}}_u^\text{H}\mathbf{\hat{v}}_i|^2\right] $ depends on the quantization error \\$\text{E}\left[ \sin^2(\measuredangle(\tilde{\mathbf{h}}_u,\hat{\mathbf{h}}_u))\right] $, i.e., for any $k\neq i$,
\begin{align} \label{eq_hkHvi}
\text{E}\left[|\tilde{\mathbf{h}}_u^\text{H}\mathbf{\hat{v}}_i|^2\right] =\alpha\text{E}\left[\sin^2(\measuredangle(\tilde{\mathbf{h}}_u,\hat{\mathbf{h}}_u))\right],
\end{align}
where the scale factor $\alpha$ can be upper bounded in the following \textbf{Lemma 2}.
\begin{IEEEproof}
$\mathbf{\tilde{h}}_u=\frac{\mathbf{h}_u}{\|\mathbf{h}_u\|}$ is the normalized channel vector of the $u$-th user.
$\hat{\mathbf{h}}_u=\|\mathbf{h}_u\|\mathbf{c}_{u,i_u}$ is the quantized channel vector,
where $\mathbf{c}_{u,i_u}$ is the unit-norm quantization vector with index $i_u$ in the codebook $\mathcal{C}_u$.
We define $Z=\sin^2(\measuredangle(\tilde{\mathbf{h}}_u,\hat{\mathbf{h}}_u))=\sin^2(\measuredangle(\tilde{\mathbf{h}}_u,\mathbf{c}_{u,i_u})$.
Since the size of the codebook is limited, it is clear that $Z\neq 0$ with probability 1.
Thus, the normalized channel vector $\mathbf{\tilde{h}}_u$ can be decomposed along two orthogonal directions,
one is the direction of quantization vector $\mathbf{c}_{u,i_u}$, and the other one is in the nullspace of $\mathbf{c}_{u,i_u}$ \cite{TIT_NJindal_MIMOBroadcast}:
 \begin{align}\label{eq12}
\tilde{\mathbf{h}}_u=\sqrt{1-Z}\mathbf{c}_{u,i_u}+\sqrt{Z}\mathbf{q},
\end{align}
where $\mathbf{q}$ is an unit-norm vector distributed in the null space of $\mathbf{c}_{u,i_u}$.
Utilizing the orthogonality between $\mathbf{c}_{u,i_u}$ and $\mathbf{q}$, we have
\begin{align}
|\tilde{\mathbf{h}}_u^\text{H}\mathbf{\hat{v}}_i|^2=(1-Z)|\mathbf{c}_{u,i_u}^\text{H}\mathbf{\hat{v}}_i|^2+Z|\mathbf{q}^\text{H}\mathbf{\hat{v}}_i|^2.
\end{align}
Since $\mathbf{\hat{v}}_i$ is the ZF precoding vector, which is orthogonal to $\mathbf{\hat{h}}_u=\|\mathbf{h}_u\|\mathbf{c}_{u,i_u}$,
we have
\begin{align}
|\tilde{\mathbf{h}}_u^\text{H}\mathbf{\hat{v}}_i|^2=Z|\mathbf{q}^\text{H}\mathbf{\hat{v}}_i|^2.
\end{align}
According to the results of \cite{TIT_NJindal_MIMOBroadcast}, $Z$ is independent with $\mathbf{q}$ and $\mathbf{\hat{v}}_i$, so we have
\begin{align}
\text{E}\left[|\tilde{\mathbf{h}}_u^\text{H}\mathbf{\hat{v}}_i|^2\right] &=\text{E}\left[Z\right] \text{E}\left[|\mathbf{q}^\text{H}\mathbf{\hat{v}}_i|^2\right] .
\end{align}
By denoting $\alpha=\text{E}\left[|\mathbf{q}^\text{H}\mathbf{\hat{v}}_i|^2\right] $, we can obtain (\ref{eq_hkHvi}).
\end{IEEEproof}

\textbf{Lemma 2}:
The scale factor can be upper bounded as $\alpha\leq\frac{1}{P-1}$ when $\hat{\mathbf{A}}_u=\mathbf{A}_u$ in the large-dimensional regime, i.e., $M\rightarrow \infty$.

\begin{IEEEproof}
Firstly, we consider the extreme case that the channels of all the $U$ users are strongly correlated. In this case, the maximum of the scale factor $\alpha$ should be achieved to ensure the largest inter-user interference $\text{E}\left[|\tilde{\mathbf{h}}_u^\text{H}\mathbf{\hat{v}}_i|^2\right]$ in (\ref{eq_hkHvi}).
Base on the ray-based channel model in Subsection \ref{S2.1}, the channels of $U$ users are strongly correlated when they share the same clusters around the BS, i.e., $P_1=P_2=\cdots=P_U=P$ and $\mathbf{A}_1 =\mathbf{A}_2=\cdots=\mathbf{A}_U=\mathbf{A}$.
Thus, we omit the subscript $u$ of $P_u$, $\mathbf{A}_u$ and $\hat{\mathbf{A}}_u$ for this subsection.
Then we look at the two vectors $\mathbf{q}$ and $\mathbf{\hat{v}}_i$, respectively.
When $\hat{\mathbf{A}}_u=\mathbf{A}_u$, i.e., $\hat{\mathbf{A}}=\mathbf{A}$, as we presented in Section \ref{S3.2},
both the fed back channel vector $\hat{\mathbf{h}}_u$ and normalized channel vector $\tilde{\mathbf{h}}_u$ are distributed in the column-space of $\mathbf{A}$.
Since $\tilde{\mathbf{h}}_u$ can be orthogonally discomposed along the directions of $\hat{\mathbf{h}}_u$ and $\mathbf{q}$,
$\mathbf{q}$ should also be distributed in the column-space of $\mathbf{A}$.
Thus, $\mathbf{q}$ can be expressed as $\mathbf{q}=\frac{\mathbf{A}\mathbf{t}}{\|\mathbf{A}\mathbf{t}\|}$,
where $\mathbf{t}\in\mathbb{C}^{P \times 1}$ and we can assume that $\|\mathbf{t}\|=1$ without loss of generality.
Similar to the proof in (\ref{eq_ci_norm}), utilizing the orthogonality among column vectors of $\mathbf{A}$ when $M\rightarrow\infty$,  we have $\|\mathbf{A}\mathbf{t}\|\overset{M\rightarrow\infty}{=}\|\mathbf{t}\|=1$. 
Therefore, $\mathbf{q}$ can be expressed as $\mathbf{q}\overset{M\rightarrow\infty}{=}\mathbf{A}\mathbf{t}$.
The precoding vector $\mathbf{\hat{v}}_i$ is the normalized $i$-th column of $\mathbf{H}^\dagger=\mathbf{H}(\mathbf{H}^\text{H}\mathbf{H})^{-1}$.
By using $\mathbf{H} = [\mathbf{h}_1,\mathbf{h}_2,\cdots,\mathbf{h}_U] = \mathbf{A}[\mathbf{g}_{1},\mathbf{g}_{2},\cdots,\mathbf{g}_{P}] =\mathbf{A}\mathbf{G}$,
we have $\mathbf{\hat{v}}_i = \frac{\mathbf{A}\mathbf{u}}{\|\mathbf{A}\mathbf{u}\|}$ where $\mathbf{u}$ is assumed as a unit-norm vector without loss of generality.
Similar to the proof in (\ref{eq_ci_norm}), utilizing the orthogonality among column vectors of $\mathbf{A}$ when $M\rightarrow\infty$,  we have $\|\mathbf{A}\mathbf{u}\|\overset{M\rightarrow\infty}{=}\|\mathbf{u}\|=1$. Therefore, $\mathbf{\hat{v}}_i$ can be expressed as $\mathbf{\hat{v}}_i \overset{M\rightarrow\infty}{=} \mathbf{A}\mathbf{u}$.
Then we calculate
\begin{align}\label{eq_qHvi}
\text{E}\left[|\mathbf{q}^\text{H}\mathbf{\hat{v}}_i|^2\right] &\overset{M\rightarrow\infty}{=}\text{E}\left[|\mathbf{t}^\text{H}\mathbf{A}^\text{H}\mathbf{A}\mathbf{u}|^2\right]  \\\nonumber
&\overset{M\rightarrow\infty}{=}\text{E}\left[|\mathbf{t}^\text{H}\mathbf{u}|^2\right] .
\end{align}
where the second equation follows from the result $\mathbf{A}^\text{H}\mathbf{A}\overset{M\rightarrow\infty}{=} \mathbf{I}_P$ in \textbf{Lemma 3} (see Appendix I).
We can also prove \textbf{Lemma 4} in Appendix II that
\begin{align}\label{eqtwH}
\text{E}\left[|\mathbf{t}^\text{H}\mathbf{u}|^2\right] =\frac{1}{P-1}.
\end{align}
By substituting (\ref{eqtwH}) into (\ref{eq_qHvi}), we have
\begin{align}\label{eqalpha}
\text{E}\left[|\mathbf{q}^\text{H}\mathbf{\hat{v}}_i|^2\right] \overset{M\rightarrow\infty}{=}\frac{1}{P-1}.
\end{align}
Note that (\ref{eqalpha}) is obtained in the extreme case that all $U$ users share the same clusters, i.e., channels of $U$ users are strongly correlated.
Therefore, the scale factor $\alpha = \text{E}\left[|\mathbf{q}^\text{H}\mathbf{\hat{v}}_i|^2\right]$ should be smaller than $\frac{1}{P-1} $.
\end{IEEEproof}

\subsection{Quantization Error in The Large-Dimensional Regime}\label{S4.2}
In this subsection, we compute the quantization error $\text{E}\left[\sin^2(\measuredangle(\tilde{\mathbf{h}}_u,\hat{\mathbf{h}}_u))\right] $ in (\ref{eq_hkHvi}) in the large-dimensional regime when the proposed AoD-adaptive subspace codebook is considered.
For the rest of this section, we omit the subscript $u$ for simplicity.
Since $\|\tilde{\mathbf{h}}\|=1$ and $\frac{\hat{\mathbf{h}}}{\|\hat{\mathbf{h}}\|} = \mathbf{c}_i$,
the quantization error $\text{E}\left[\sin^2(\measuredangle(\tilde{\mathbf{h}},\hat{\mathbf{h}}))\right] $ can be expressed as
\begin{align}\label{sin2h}
\text{E}\left[\sin^2(\measuredangle(\tilde{\mathbf{h}},\hat{\mathbf{h}}))\right] =1-\text{E}\left[|\tilde{\mathbf{h}}^\text{H}\mathbf{c}_i|^2\right] ,
\end{align}
where $\tilde{\mathbf{h}}=\frac{\mathbf{A}\mathbf{g}}{\|\mathbf{h}\|}$ according to (\ref{eqhk2}).
Similar to the proof in (\ref{eq_ci_norm}), it is easy to show that $\|\mathbf{h}\|=\|\mathbf{A}\mathbf{g}\|\overset{M\rightarrow\infty}{=}\|\mathbf{g}\|$.
By denoting $\tilde{\mathbf{g}}=\frac{\mathbf{g}}{\|\mathbf{g}\|}$,
we have $\tilde{\mathbf{h}}\overset{M\rightarrow\infty}{=}\frac{\mathbf{A}\mathbf{g}}{\|\mathbf{g}\|}\overset{M\rightarrow\infty}{=}\mathbf{A}\tilde{\mathbf{g}}$.
Combining $\tilde{\mathbf{h}}\overset{M\rightarrow\infty}{=}\mathbf{A}\tilde{\mathbf{g}}$ and (\ref{eq_cki}), we have
\begin{align}\label{hc}
\text{E}\left[|\tilde{\mathbf{h}}^\text{H}\mathbf{c}_i|^2\right] &\overset{M\rightarrow\infty}{=}\text{E}\left[|\tilde{\mathbf{g}}^\text{H}\mathbf{A}^\text{H}\hat{\mathbf{A}}\mathbf{w}_i|^2\right] \\\nonumber
&\overset{M\rightarrow\infty}{=}\text{E}\left[|\mathcal{K}\tilde{\mathbf{g}}^\text{H}\mathbf{w}_i|^2\right] \\\nonumber
&\overset{M\rightarrow\infty}{=}|\mathcal{K}|^2\text{E}\left[|\tilde{\mathbf{g}}^\text{H}\mathbf{w}_i|^2\right] ,
\end{align}
where the second equation is true due to $\mathbf{A}^\text{H}\hat{\mathbf{A}}\overset{M\rightarrow\infty}{=} \mathcal{K} \mathbf{I}_P$ as proved in \textbf{Lemma 3} (see Appendix I).
Since both $\tilde{\mathbf{g}}$ and $\mathbf{w}_i$ are isotropically distributed vectors on the $P$-dimensional unit sphere, we have \cite{TIT_NJindal_MIMOBroadcast}
\begin{align}\label{gupbound}
\text{E}\left[|\tilde{\mathbf{g}}^\text{H}\mathbf{w}_i|^2\right] >1-2^{-\frac{B}{P-1}}.
\end{align}
Combining (\ref{sin2h}), (\ref{hc}), and (\ref{gupbound}), we have
\begin{align} \label{eq_Qua_err1}
\text{E}\left[\sin^2(\measuredangle(\tilde{\mathbf{h}},\hat{\mathbf{h}}))\right] <1-\left|\mathcal{K}\right|^2(1-2^{-\frac{B}{P-1}}).
\end{align}
By substituting $\left|\mathcal{K}\right|^2\geq 1-\frac{M^2}{3}\left(\pi\frac{d}{\lambda}\right)^2r^22^{-2B_0}$ (\ref{eq_U_M}) in Appendix I into (\ref{eq_Qua_err1}) and denoting $\beta=\frac{M^2}{3}(\pi\frac{d}{\lambda})^2r^22^{-2B_0}$,
we can obtain the upper bound of the quantization error $\text{E}\left[\sin^2(\measuredangle(\tilde{\mathbf{h}},\hat{\mathbf{h}}))\right] $ as
\begin{align} \label{eq_Qua_err2}
\text{E}\left[\sin^2(\measuredangle(\tilde{\mathbf{h}},\hat{\mathbf{h}}))\right] <\beta(1-2^{-\frac{B}{P-1}})+2^{-\frac{B}{P-1}}.
\end{align}
We can observe from (\ref{eq_Qua_err2}) that a small $\beta$ leads to a small quantization error due to $2^{-\frac{B}{P-1}}\ll 1$, where a small $\beta$ can be achieved with a large number of AoD quantization  bits $B_0$. In addition, (\ref{eq_Qua_err2}) can be rewrite as
\begin{align}
\text{E}\left[\sin^2(\measuredangle(\tilde{\mathbf{h}},\hat{\mathbf{h}}))\right] <(1-\beta)2^{-\frac{B}{P-1}}+\beta,
\end{align}
where a small quantization error can be achieved with a large number of feedback bits $B$ since $\beta\ll 1$ with a larger number of AoD quantization  bits $B_0$.

\subsection{Feedback Bits}\label{S4.3}
Finally, based on the previous analytical results in the large-dimensional regime, we will evaluate the required number of feedback bits $B$ to guarantee a constant rate gap $\Delta R_\text{Quantized}$.
By combining (\ref{delatR1}), (\ref{eq_hkHvi}), and (\ref{eq_Qua_err2}), we can obtain the rate gap as
\begin{eqnarray}\label{delatR3}
\Delta R_\text{Quantized}\leq &\log_2\left(1+\frac{(U-1)\gamma}{U} \text{E}\left[\|\mathbf{h}\|^2\right] \alpha 2^{-\frac{B}{P-1}}+\frac{(U-1)\gamma}{U}\text{E}\left[\|\mathbf{h}\|^2\right] \alpha \beta(1-2^{-\frac{B}{P-1}})\right).
\end{eqnarray}
If we consider the perfect case of AoD acquisition, i.e., $B_0$ is large enough, we have $\beta=0$,
and (\ref{delatR3}) can be simplified as
\begin{align}\label{delatR4}
\Delta R_\text{Quantized}&\leq \log_2\left(1+\frac{(U-1)\gamma}{U}\text{E}\left[\|\mathbf{h}\|^2\right] \alpha 2^{-\frac{B}{P-1}}\right).
\end{align}
Let $\Delta R_\text{Quantized}\leq \log_2\left(1+\frac{(U-1)\gamma}{U}\text{E}\left[\|\mathbf{h}\|^2\right] \alpha 2^{-\frac{B}{P-1}}\right) \leq \log_2(b)$ bps/Hz, then the number of feedback bits $B$ should scale according to
\begin{align}\label{eq_B}
B \geq \frac{P-1}{3}\text{SNR}+(P-1)\log_2{\frac{(U-1)\alpha}{b-1}},
\end{align}
where $\text{SNR}=10\log_{10}\frac{\gamma}{U}\text{E}\left[\|\mathbf{h}\|^2\right] $ is the signal-to-noise-ratio (SNR) at the receiver.

\textit{\textbf{Remark 1}}:
We observe that the slope of the required number of feedback bits $B$ is $P-1$ when $\text{SNR}$ increases.
In other words, the required number of feedback bits only scales linearly with $P-1$ to maintain a constant rate gap.
Since $P\ll M$,
the proposed AoD-adaptive subspace codebook can reduce the codebook size and feedback overhead significantly.

\section{Quantized Channel Feedback vs. Analog Channel Feedback}\label{S5}
Analog linear modulation may be used for channel state feedback without quantization (i.e., analog channel feedback)  due to the circumvention of codebook complexities and delays associated with quantization, source coding and channel coding for quantized channel feedback \cite{TSP_TLMarzetta_FastTransfer}.
Previous work has compared analog channel feedback with quantized channel feedback using RVQ-based codebook for quantization \cite{ISIT_GCaire_QuantizedVSAnalog}. It has been shown that analog channel feedback can not be outperformed by quantized channel feedback using RVQ-based codebook for quantization since the analog transmission is optimal when a Gaussian channel vector is transmitted over a noisy AWGN channel.
If the channel rate (of uplink) is larger than the source rate (i.e., downlink CSI), quantized channel feedback using RVQ-based codebook for quantization is superior to analog channel feedback because the effect of feedback noise vanishes at high SNR for quantized channel feedback but does not do so for analog. In this section, we will compare analog channel feedback and quantized channel feedback when the proposed AoD-adaptive subspace codebook is used. 

We first analyze the sum rate in the large-dimensional regime when a novel analog channel feedback scheme is utilized to obtain CSI at the BS. Consider the same downlink massive MIMO systems in section \ref{S2} with $M$ BS antennas and $U$ single-antenna users, where ZF precoding is utilized to eliminate multi-user interference.
Different from the aforementioned quantized channel feedback through digital uplink channel, the CSI at the BS is obtained through the analog noisy uplink channel. Specifically, the elements of channel vector $\mathbf{h}_u$ in (\ref{eqhk1}) are explicitly fed back through the uplink noisy channel with unquantized quadrature-amplitude modulation in traditional analog channel feedback schemes \cite{TSP_TLMarzetta_FastTransfer,TCOM_HShirani-Mehr_CSIFeedbackMUMIMO}.
In this paper, we assume that path AoDs, i.e., steering matrix $\mathbf{A}_u$ of the $u$-th user is known to the BS and the $u$-th user during the angle coherence time for both quantized channel feedback using the proposed AoD-adaptive subspace codebook and analog channel feedback for fair comparison.
In this case, with $\mathbf{h}_u=\mathbf{A}_u\mathbf{g}_u$, only the elements of the $P_u\times 1$ path gain vector $\mathbf{g}_u$ are fed back through noisy analog uplink channel for CSI feedback, which is different from the traditional analog channel feedback scheme where all the elements of $M\times 1$ channel vector $\mathbf{h}_u$ of the $u$-th user are fed back without quantization. We assume no fading on the uplink channel for simplicity and the observation of path gain vector $\mathbf{z}_u$ of the $u$-th user at the BS after analog uplink channel is given by \cite{ISIT_GCaire_QuantizedVSAnalog}
\begin{align}\label{eq_z}
\mathbf{z}_u=\sqrt{\mu\gamma_\text{U}}\mathbf{g}_u+\mathbf{n}_{\text{U},u},
\end{align}
where $\gamma_\text{U}$ is the uplink SNR and $\mathbf{n}_{\text{U},u}$ is the uplink complex Gaussian noise of the $u$-th user, where each elements have zero mean and unit variance. The scaling factor $\mu$ denotes the number of channel uses per element of path gain vector $\mathbf{g}_u$. This equation (\ref{eq_z}) models the case where the elements of path gain vector $\mathbf{g}_u$ are fed back through uplink channel without quantization. The MMSE estimate of the path gain vector at the BS is $\check{\mathbf{g}}_u=\frac{\sqrt{\mu\gamma_\text{U}}}{1+\mu\gamma_\text{U}}\mathbf{z}_u$. We denote $\mathbf{g}_u = \check{\mathbf{g}}_u + \mathbf{e}_{g,u}$, where $\check{\mathbf{g}}_u$ and $\mathbf{e}_{g,u}$ are mutually independent and have Gaussian components with zero mean and variance $\mu\gamma_\text{U}\sigma_{e_{g,u}}^2$ and $\sigma_{e_{g,u}}^2=(1+\mu\gamma_\text{U})^{-1}$.
Then, by utilizing the path AoD information, i.e., steering matrix $\mathbf{A}_u$, the BS can obtain the channel vector $\check{\mathbf{h}}_u$ obtained from analog feedback as 
\begin{align}\label{eq_hcheck}
\check{\mathbf{h}}_u=\mathbf{A}_u\check{\mathbf{g}}_u.
\end{align}
Therefore we can rewrite the channel vector $\mathbf{h}_u$ as 
\begin{align}\label{eq_hkanalog}
\mathbf{h}_u=\mathbf{A}_u\mathbf{g}_u = \check{\mathbf{h}}_u + \mathbf{A}_u\mathbf{e}_{g,u}.
\end{align}
The concatenation of the channel vectors obtained from analog feedback can be denoted as $\mathbf{\check{H}}=[\mathbf{\check{h}}_1,\mathbf{\check{h}}_2,\cdots,\mathbf{\check{h}}_U] \in\mathbb{C}^{M\times U}$.

Similar to quantized channel feedback, we consider the ZF precoding before downlink transmission, which is realized based on the channel matrix $\mathbf{\check{H}}$ obtained from analog channel  feedback. The ZF precoding matrix $\mathbf{\check{V}}=[\mathbf{\check{v}}_1,\mathbf{\check{v}}_2,\cdots,\mathbf{\check{v}}_U]\in\mathbb{C}^{M\times U}$ consists of $U$ different $M$-dimensional unit-norm precoding vectors $\mathbf{\check{v}}_i\in\mathbb{C}^{M\times 1}$,
 which is obtained as the normalized $i$-th column of ${\mathbf{\check{H}}}^\dagger$, i.e., $\mathbf{\check{v}}_i=\frac{{\mathbf{\check{H}}}^\dagger(:,i)}{\|{\mathbf{\check{H}}}^\dagger(:,i)\|}$.
 After the downlink channel, the received signal at the $u$-th user can be described as
\begin{align}
y_u&=\sqrt{\frac{\gamma}{U}}\mathbf{h}_u^\text{H}\mathbf{\check{v}}_us_u+\sqrt{\frac{\gamma}{U}}\sum_{i=1,i\neq u}^{U}\mathbf{h}_u^\text{H}\mathbf{\check{v}}_is_i+n_u \\\nonumber
&\overset{(a)}{=}\sqrt{\frac{\gamma}{U}}\mathbf{h}_u^\text{H}\mathbf{\check{v}}_us_u+\sqrt{\frac{\gamma}{U}}\sum_{i=1,i\neq u}^{U}\mathbf{e}_{g,u}^\text{H}\mathbf{A}_u^\text{H}\mathbf{\check{v}}_is_i+n_u,
\end{align}
where (a) is obtained by substituting (\ref{eq_hkanalog}) and $\check{\mathbf{h}}_u^\text{H}\mathbf{\check{v}}_i=0$ for $i\neq u $.
The per-user rate with analog channel feedback is
 \begin{align} \label{eq_R_Analog}
R_\text{Analog}=\text{E}\left[\log_2\left(1+\frac{\frac{\gamma}{U}|\mathbf{h}_u^\text{H}\mathbf{\check{v}}_u|^2}{1+\frac{\gamma}{U}\sum_{i=1,i\neq u}^{U}|\mathbf{e}_{g,u}^\text{H}\mathbf{A}_u^\text{H}\mathbf{\check{v}}_i|^2}\right)\right].
\end{align}
Defining the rate gap between the ideal case of perfect CSIT and the practical case of analog channel feedback as $\Delta R_\text{Analog} =R_\text{Ideal} - R_\text{Analog}$, we have 
 \begin{align}
\Delta R_\text{Analog} = \text{E}\left[\log_2\left(1+\frac{\gamma}{U}|\mathbf{h}_u^\text{H}\mathbf{v}_{\text{Ideal},i}|^2\right)\right] - \text{E}\left[\log_2\left(1+\frac{\frac{\gamma}{U}|\mathbf{h}_u^\text{H}\mathbf{\check{v}}_u|^2}{1+\frac{\gamma}{U}\sum_{i=1,i\neq u}^{U}|\mathbf{e}_{g,u}^\text{H}\mathbf{A}_u^\text{H}\mathbf{\check{v}}_i|^2}\right)\right].
\end{align}
Following \cite{ISIT_GCaire_QuantizedVSAnalog}, the rate gap with analog channel feedback $\Delta R_\text{Analog}$ can be upper bounded as 
\begin{align}\label{eq_Ranalog1}
\Delta R_\text{Analog} \leq  \log_2\left(1+(U-1)\frac{\gamma}{U}\text{E}\left[|\mathbf{e}_{g,u}^\text{H}\mathbf{A}_u^\text{H}\mathbf{\check{v}}_i|^2\right]\right).
\end{align}
Next we evaluate the inter-user interference $\text{E}\left[|\mathbf{e}_{g,u}^\text{H}\mathbf{A}_u^\text{H}\mathbf{\check{v}}_i|^2\right]$.
Consider the extreme case that the channels of all the $U$ users are strongly correlated, i.e., the $U$ users share the same clusters around the BS.
In this extreme case, the inter-user interference $\text{E}\left[|\mathbf{e}_{g,u}^\text{H}\mathbf{A}_u^\text{H}\mathbf{\check{v}}_i|^2\right]$ can achieve its upper bound.
Based on the ray-based channel model (\ref{eqhk1}), we have  $P_1=P_2=\cdots=P_U=P$ and $\mathbf{A}_1 =\mathbf{A}_2=\cdots=\mathbf{A}_U=\mathbf{A}$. Therefore, we omit the subscript $u$ of $P_u$ and $\mathbf{A}_u$ for the rest of this subsection.
Since the ZF precoding vector $\mathbf{\check{v}}_i$ is the normalized $i$-th column of $\mathbf{\check{H}}^\dagger=\mathbf{\check{H}}(\mathbf{\check{H}}^\text{H}\mathbf{\check{H}})^{-1}$, by combining (\ref{eq_hcheck}), we can rewrite that $\mathbf{\check{v}}_i = \mathbf{A}\mathbf{p}$ where $\|\mathbf{p}\| = 1$. Therefore we can upper bound the inter-use interference as 
\begin{align}\label{eq_eAv1}
\text{E}\left[|\mathbf{e}_{g,u}^\text{H}\mathbf{A}_u^\text{H}\mathbf{\check{v}}_i|^2\right] \leq \text{E}\left[|\mathbf{e}_{g,u}^\text{H}\mathbf{A}^\text{H}\mathbf{A}\mathbf{p}|^2\right] = \text{E}\left[|\mathbf{p}^\text{H}\mathbf{A}^\text{H}\mathbf{A}\text{E}\left[\mathbf{e}_{g,u}\mathbf{e}_{g,u}^\text{H}\right]\mathbf{A}^\text{H}\mathbf{A}\mathbf{p}|\right]
\end{align}
By substituting $\mathbf{A}^\text{H}\mathbf{A}\overset{M\rightarrow\infty}{=}\mathbf{I}_P$ and $\text{E}\left[\mathbf{e}_{g,u}\mathbf{e}_{g,u}^\text{H}\right] = \sigma_{e_{g,u}}^2\mathbf{I}_P$, we have 
\begin{align}\label{eq_eAv2}
\text{E}\left[|\mathbf{e}_{g,u}^\text{H}\mathbf{A}_u^\text{H}\mathbf{\check{v}}_i|^2\right] \leq \sigma_{e_{g,u}}^2.
\end{align}
By combining (\ref{eq_Ranalog1}) and (\ref{eq_eAv2}), we obtain the upper bound of the rate gap with analog channel feedback
\begin{align}\label{eq_Ranalog2}
\Delta R_\text{Analog} \leq  \log_2\left(1+(U-1)\frac{\gamma}{U}\sigma_{e_{g,u}}^2\right)=\log_2\left(1+(U-1)\frac{\gamma}{U}(1+\mu\gamma_\text{U})^{-1}\right).
\end{align}

Now we recall the rate gap with quantized channel feedback using the proposed AoD-adaptive subspace codebook. Assuming that AoDs are perfectly known, the rate gap $\Delta R_\text{Quanized}$ with quantized feedback using the proposed AoD-adaptive subspace codebook is expressed in (\ref{delatR4}).
Similar to (\ref{eq_ci_norm}), we have $\text{E}\left[\|\mathbf{h}\|^2\right] =P$. Therefore the rate gap $\Delta R_\text{Quanized}$ with quantized channel feedback can be further bounded as 
 \begin{align}\label{delatR6}
 \Delta R_\text{Quanized} \leq\log_2\left(1+\gamma \frac{P}{P-1} 2^{-\frac{B}{P-1}}\right).
 \end{align}
Following the assumption in \cite{ISIT_GCaire_QuantizedVSAnalog} that the digital feedback link can be operated error-free and at capacity, i.e., $\log_2(1+\gamma_\text{U})$ bits can be transmitted per symbol transmission. For the analog feedback of path gain vector $\mathbf{g}$, $\mu P$ symbols are transmitted.  $B=\mu P \log_2(1+\gamma_\text{U})$ bits can be transmitted using the same feedback resource.
By substituting $B=\mu P \log_2(1+\gamma_\text{U})$ into (\ref{delatR6}), we obtain the rate gap $\Delta R_\text{Quanized}$ with quantized feedback using the proposed AoD-adaptive subspace codebook
 \begin{align}\label{delatR7}
\Delta R_\text{Quanized} \leq\log_2\left(1+\gamma \frac{P}{P-1} {(1+\gamma_\text{U})}^{-\frac{\mu P}{P-1}}\right). 
\end{align}

 \textit{\textbf{Remark 2}}: We consider the case of constant uplink SNR $\gamma_\text{U}$. For analog channel feedback (\ref{eq_Ranalog2}), $2^{\Delta R_\text{Analog}}$ decreases inversely with the scale factor $\mu$. While for quantized channel feedback using the proposed AoD-adaptive subspace codebook (\ref{delatR7}), $2^{\Delta R_\text{Quanized}}$ decreases exponentially with the scale factor $\mu$. Therefore, we can conclude that quantized channel feedback outperforms analog channel feedback when $\mu$ is large. This result is consistent with the previous conclusion in \cite{ISIT_GCaire_QuantizedVSAnalog} that quantized channel feedback is superior to analog channel feedback when the channel rate is larger than the source rate (i.e., $\mu$ is large), which is obtained when the RVQ-based codebook is utilized for quantization.
 
  \textit{\textbf{Remark 3}}: Consider the case of constant scale factor $\mu$. For analog channel feedback (\ref{eq_Ranalog2}), $2^{\Delta R_\text{Analog}}$ decreases inversely with the uplink SNR $\gamma_\text{U}$. For quantized channel feedback using the proposed AoD-adaptive subspace codebook (\ref{delatR7}), $2^{\Delta R_\text{Quanized}}$ decreases proportionally with $\frac{\mu P}{P-1}$-power of $\gamma_\text{U}$.  
  Therefore, analog channel feedback outperforms quantized channel feedback using the proposed AoD-adaptive subspace codebook when $\mu < \dfrac{P-1}{P}$. This is consistent with classical conclusion that the analog transmission is optimal when a Gaussian source is transmitted over a AWGN channel \cite{TIT_MGastpar_CodeorNot}.
  However, when the uplink channel rate is larger than the source rate, i.e., $\mu > \dfrac{P-1}{P}$ in our case, quantized channel feedback using the proposed AoD-adaptive subspace codebook outperforms analog channel feedback for high uplink SNR.
  
\section{Simulation Results}\label{S6}
 A simulation study was carried out to verify the performance of the proposed
 AoD-adaptive subspace codebook.
 The main system parameters are described in Table \ref{Table_1}.
 We consider the UPA for BS antenna with array response $\mathbf{a}(\phi_{u,i},\theta_{u,i})$ in (\ref{eq_ste_vec_UPA}).
 We assume that the number of paths is the same, i.e., $P=4$ for each user.
 The per-user rate of ideal case with perfect CSIT $R_\text{Ideal}$ is calculated according to the equation (\ref{eq_R_ideal}).
 The per-user rate of the practical case with quantized channel feedback $R_\text{Quantized}$ is calculated according to the equation (\ref{eq_R_quantized}).  
 The per-user rate of the practical case with analog channel feedback $R_\text{Analog}$ is calculated according to the equation (\ref{eq_R_Analog}), where AoDs are assumed to be known and only the path gain vector $\mathbf{g}_k$ are fed back through noisy uplink channel as shown in (\ref{eq_z}). 
 We assume that AoDs are perfectly known to the BS and users for the proposed AoD-adaptive subspace codebook except for Fig. \ref{Fig6} and Fig. \ref{Fig7}.
 For fair comparison, we also assume that the transmit channel correlation matrix is perfectly known for the conventional channel statistics-based codebook. 
 The proposed AoD-adaptive subspace codebook is generated under RVQ framework except for Fig. \ref{Fig5}, i.e., $\mathbf{w}_i$ in (\ref{eq_cki}) is chosen from RVQ-based codebook. 
\begin{table}[tb!]
	\begin{center}
		\caption{System parameters for simulation}  \label{Table_1}
		\begin{tabular}{|c|c|}
			\hline
			\tabincell{c}{The number of BS antennas $M$ } & 128 \\
			\hline
			\tabincell{c}{The number of users $U$} & 4 \\
	        \hline
			\tabincell{c}{The number of resolvable paths $P$ } & 4 \\
			\hline
			\tabincell{c}{The azimuth AoD $\phi$ and the elevation AoD $\theta$} & Uniform distribution $\mathcal{U}[-\frac{1}{2}\pi, \frac{1}{2}\pi]$ \\
			\hline
			\tabincell{c}{ The receiver SNR in  (\ref{eq_B})} &  0 dB-12 dB \\
			\hline
		\end{tabular}
	\end{center}
\end{table}

\begin{figure}[tb!]
	\vspace*{-1mm}
	\begin{center}
		\includegraphics[width=0.7\textwidth]{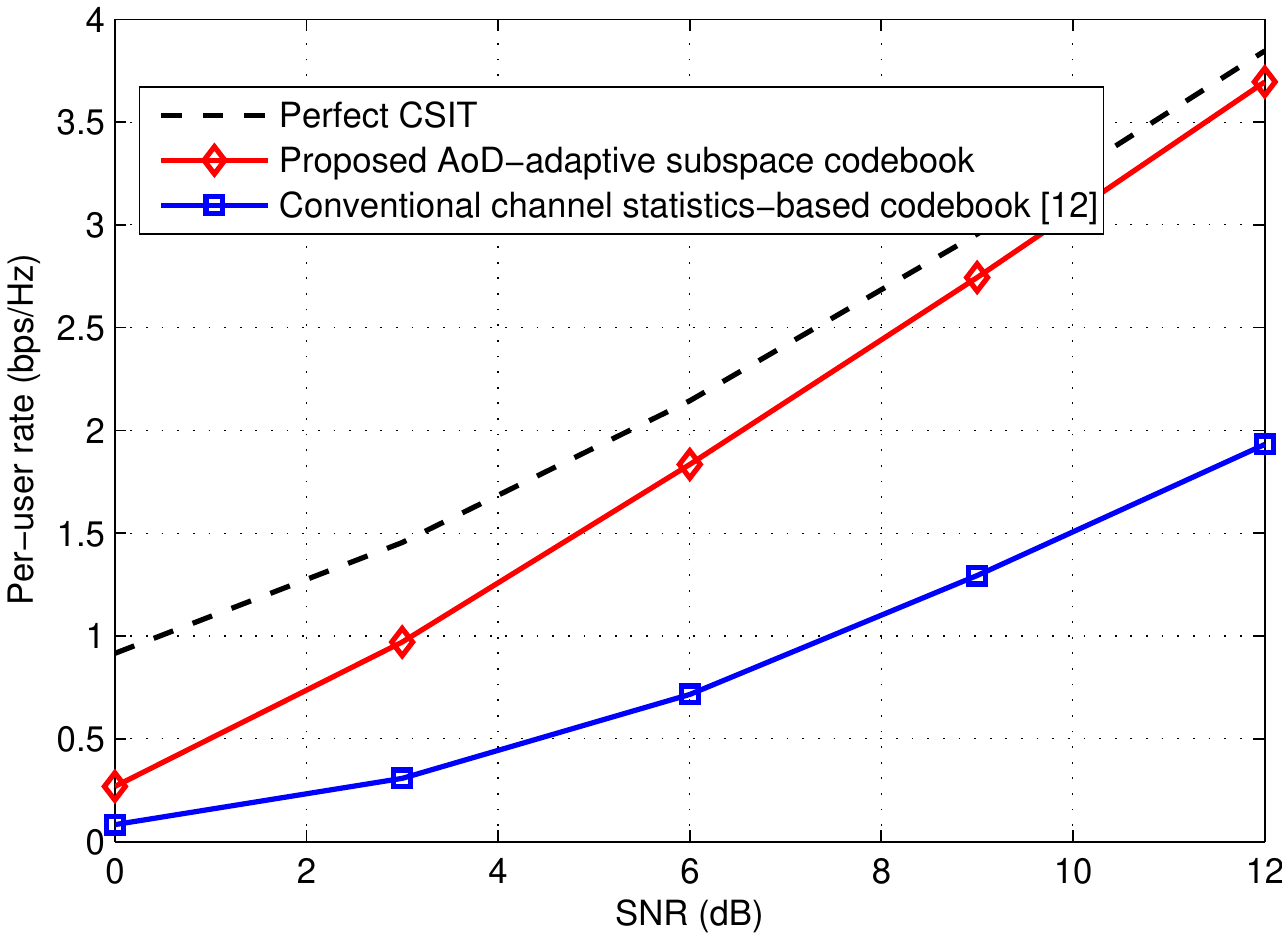}
	\end{center}
	\vspace*{-4mm}
	\caption{The per-user rate of the ideal case with perfect CSIT and the practical cases with quantized channel feedback using the proposed codebook, the conventional channel statistics-based codebook. The number of feedback bits is set as $B=\frac{P-1}{3}\text{SNR}$ for all quantized channel feedback schemes.}
	\label{Fig3}
\end{figure}

Fig.~\ref{Fig3} compares the per-user rate between the ideal case of perfect CSIT and the practical cases of quantized channel feedback,
where the proposed AoD-adaptive subspace codebook and the conventional channel statistics-based codebooks\cite{TVT_DJLove_LimitedFeedback} are considered.
The number of feedback bits $B=\frac{P-1}{3}\text{SNR}$ is same for all quantized channel feedback schemes.
We observe that the rate gap between the ideal case of perfect CSIT and the practical case of using the proposed AoD-adaptive subspace codebook can be limited within a constant value when SNR at the receiver in (\ref{eq_B}) increases,
which is consistent with our theoretical analysis in Section \ref{S4.3}.
On the contrary, for the conventional channel statistics-based codebook,
the rate gap increases with the SNR.
In addition, the proposed AoD-adaptive subspace codebook outperforms the conventional channel statistics-based codebook in terms of the per-user rate.

\begin{figure} [tb!]
	\vspace{-1mm}
	\center{\includegraphics[width=0.7\textwidth]{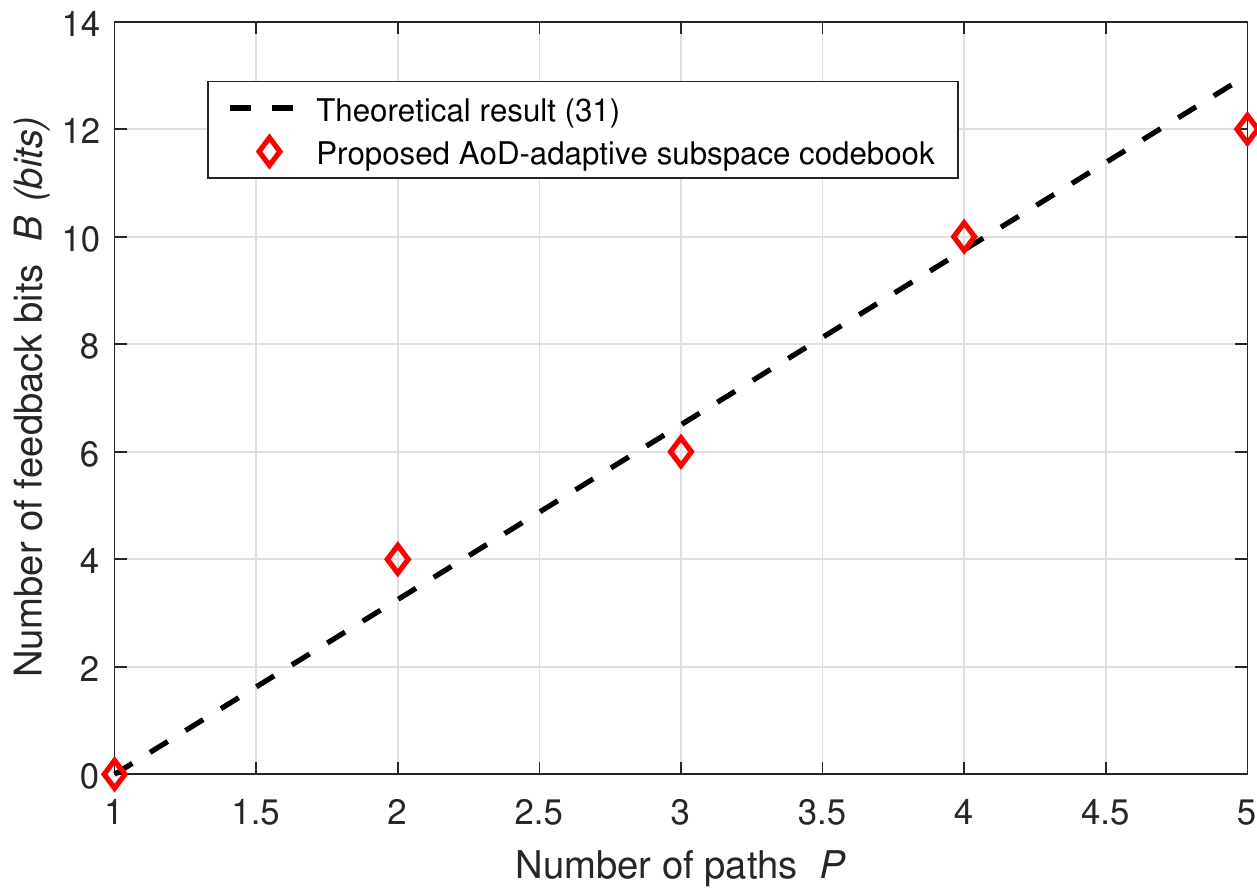}}
	\vspace{-4mm}
	\caption{The required number of feedback bits $B$ against number of paths $P$ to limit the rate gap $ \Delta R_\text{Quanized}$ within $0.13$ bps/Hz with receiver SNR=5 dB. The required number of feedback bits $B$ using the proposed AoD-adaptive subspace codebook scales linearly with the number of resolvable paths $P$.}
	\label{Fig4}
\end{figure}

\begin{figure} [tb!]
	\vspace{-1mm}
	\center{\includegraphics[width=0.7\textwidth]{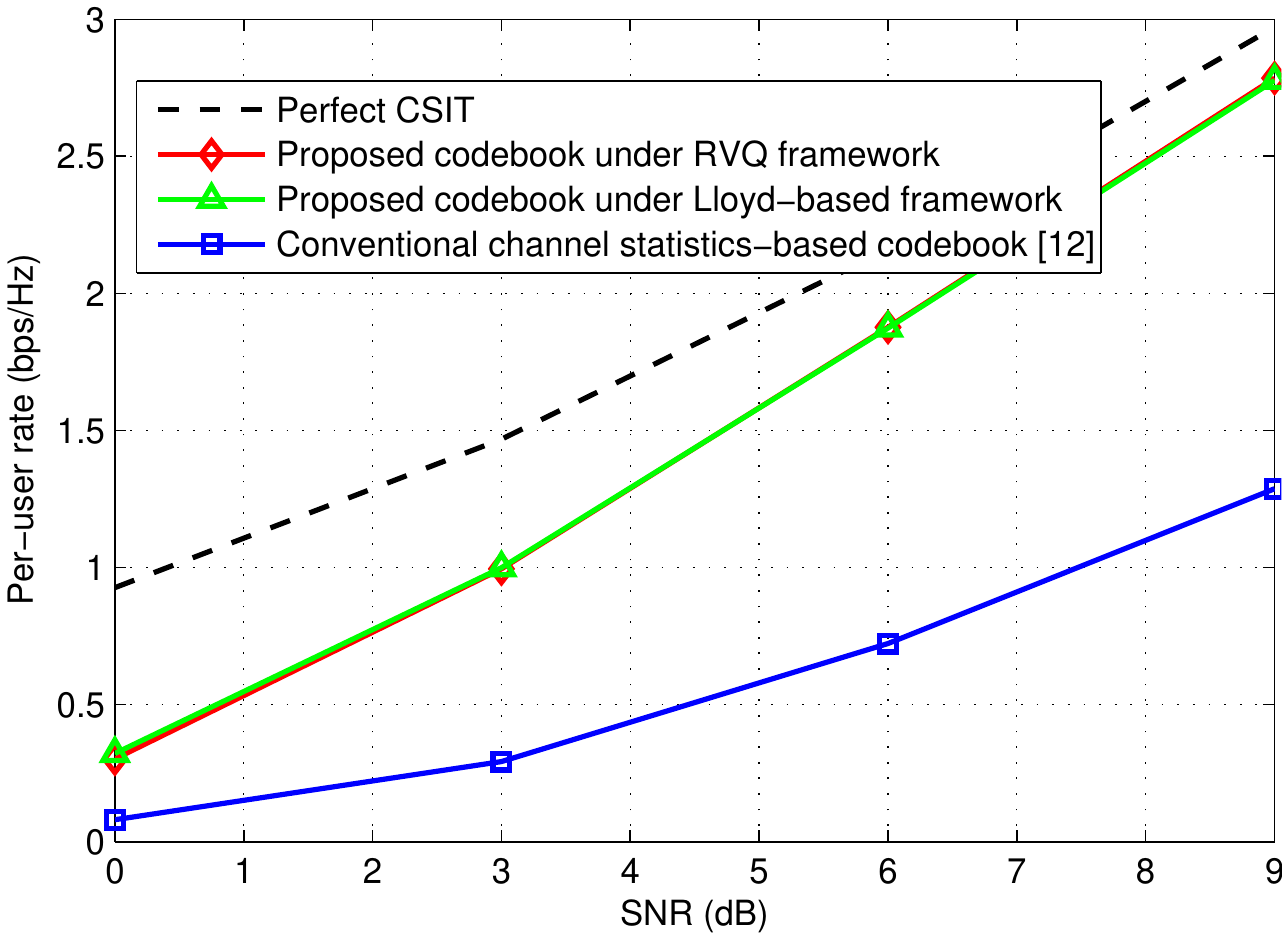}}
	\vspace{-4mm}
	\caption{The per-user rate of the ideal case with perfect CSIT and the practical cases with quantized channel feedback using the proposed codebook under RVQ framework and Lloyd-based framework, and the conventional channel statistics-based codebook. The number of feedback bits is set as $B=\frac{P-1}{3}\text{SNR}$ for all quantized channel feedback schemes.}
	\label{Fig5}
\end{figure}

\begin{figure} [tb!]
	\vspace{-1mm}
	\center{\includegraphics[width=0.7\textwidth]{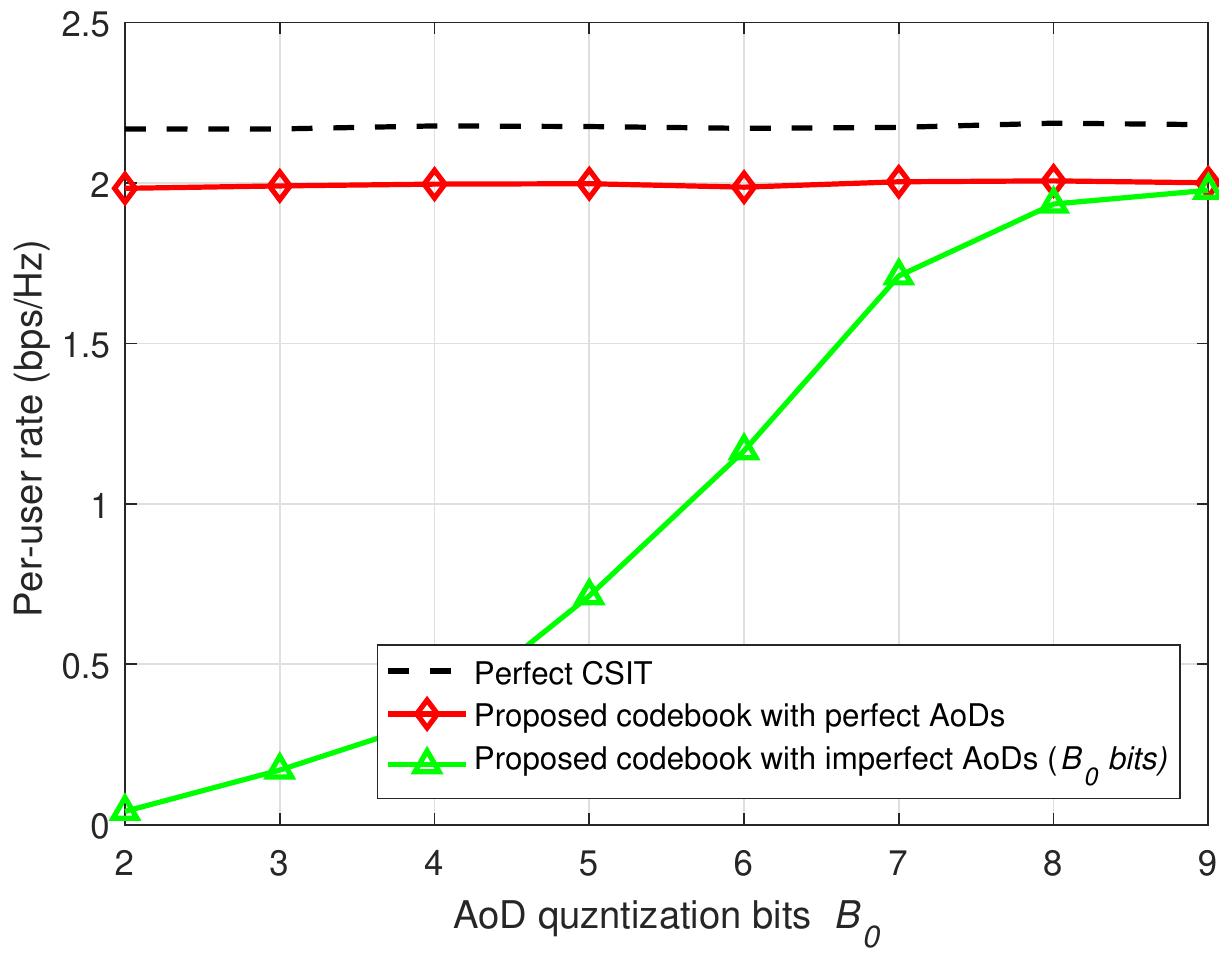}}
	\vspace{-4mm}
	\caption{The per-user rate of the proposed AoD-adaptive subspace codebook against the number of AoD quantization bits $B_0$. The receiver SNR is set as 6 dB and the number of channel feedback bits $B$ is set as 8. Since the angle coherence time is comparably long, the average required number of AoD quantization bits is very small to achieve the similar per-user rate with perfect AoDs.}
	\label{Fig6}
\end{figure}

\begin{figure}[tb!]
\vspace*{-1mm}
\begin{center}
 \includegraphics[width=0.7\columnwidth, keepaspectratio]{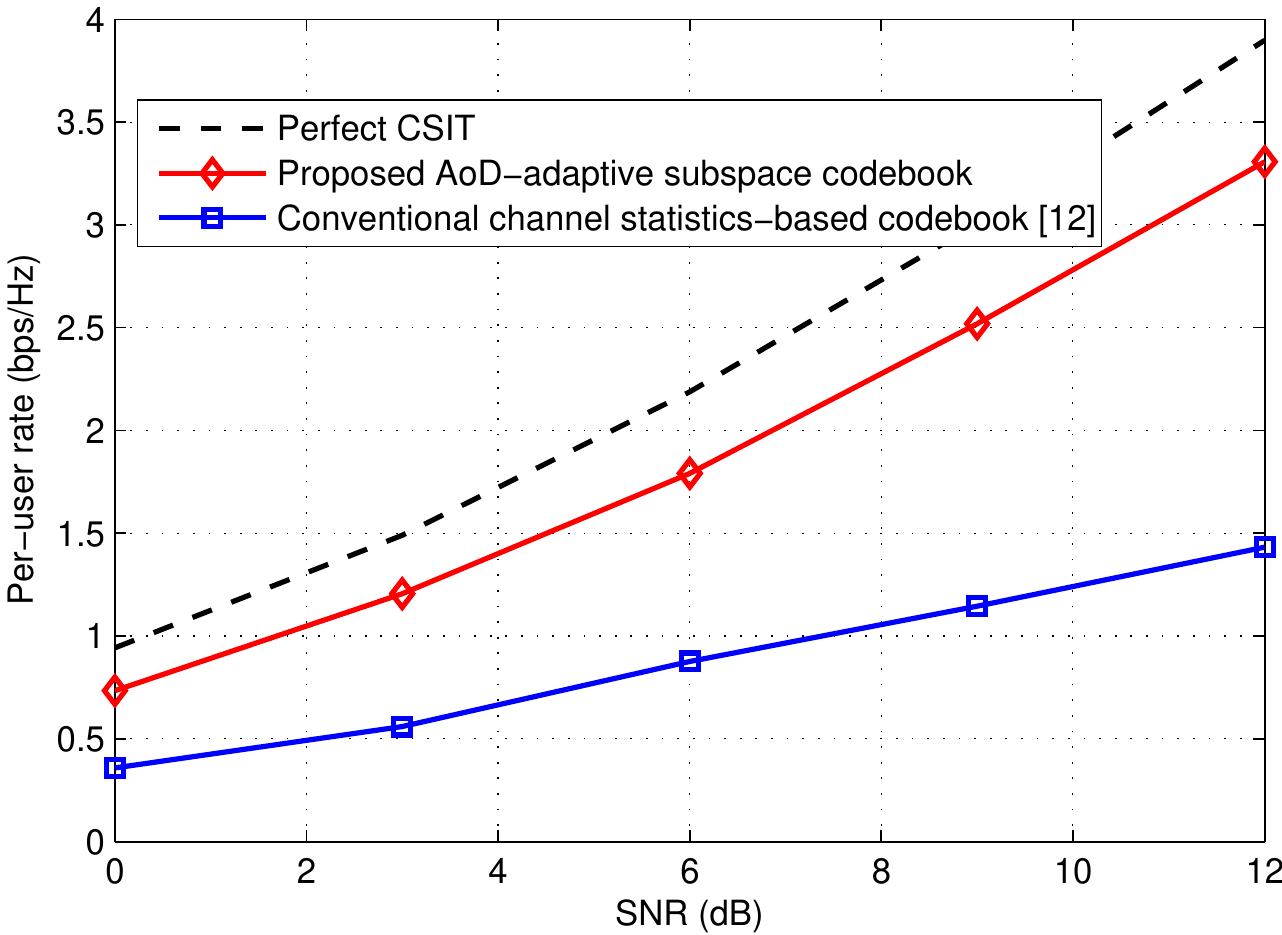}
\end{center}
\vspace*{-4mm}
 \caption{The per-user rate of the ideal case with perfect CSIT and the practical cases with quantized channel feedback, where the total number of feedback bits is set as 8 for all quantized channel feedback schemes. The codebook size of the conventional channel statistics-based codebook is $2^8=256$. For the proposed codebook, the average number of AoD quantization bits is $3$ and the number of channel feedback bits $B=8-3=5$. Therefore, the codebook size of the proposed codebook is $2^5=32$.}
 \label{Fig7}
\vspace*{-1mm}
\end{figure}

Fig. \ref{Fig4} shows the required number of feedback bits $B$ to limit the rate gap $ \Delta R_\text{Quanized}$ between the ideal case of perfect CSIT and the practical case of using the proposed AoD-adaptive subspace codebook within $0.13$ bps/Hz.
We observe that the required number of feedback bits $B$ scales linearly with the number of resolvable paths $P$.
It is consistent with the theoretical result (\ref{eq_B}) which is also shown in Fig. \ref{Fig4} for comparison.

Fig. \ref{Fig5} compares the per-user rate of the proposed AoD-adaptive subspace codebooks under RVQ framework (i.e., $\mathbf{w}_i$ in (\ref{eq_cki}) is chosen from the RVQ-based codebook) and Lloyd-based framework (i.e., $\mathbf{w}_i$ in (\ref{eq_cki}) is chosen from the optimal vector quantization codebook generated by Lloyd algorithm). We observe that the proposed AoD-adaptive subspace codebook under Lloyd-based framework slightly outperforms that under RVQ framework. Their per-user rate are very close.

In Fig. \ref{Fig6}, we show the per-user rate of the proposed AoD-adaptive subspace codebook when AoDs are not perfectly known, where each AoD is quantized with $B_0$ bits and the proposed codebook is generated based on the quantized AoDs.
 The receiver SNR is set as 6 dB and the number of channel feedback bits $B$ is set as 8.
We observe that the per-user rate of proposed AoD-adaptive subspace codebook increases with the number of AoD quantization bits $B_0$.
Note that when $B_0=8$,
the per-user rate using the proposed AoD-adaptive subspace codebook with imperfect AoDs is close to that with perfect AoDs.
Since the angle coherence time is comparably long, the average required number of AoD quantization  bits is very small.


Since the number of feedback bits is usually a constant value in practical systems, we compare the performance of the proposed AoD-adaptive subspace codebook and the conventional channel statistics-based codebook in Fig. \ref{Fig7},
where the total number of feedback bits is set as 8.
For the conventional channel statistics-based codebook,
we assume the channel correlation matrix is known to both BS and users without any additional overhead.
Thus, the codebook size is $2^8=256$.
For the proposed AoD-adaptive subspace codebook, the number of AoD quantization  bits $B_0=8$. Assuming the angle coherence time is ten times of the channel coherence time, the average number of  AoD quantization bits is $P\times B_0/10\approx3$.
Thus, the number of channel feedback bits is set as $B=8-3=5$ and the codebook size is $2^{5}=32$.
Although the codebook size of the proposed AoD-adaptive subspace codebook is much smaller than that of the classical channel statistics-based codebook,
we still observe that the proposed AoD-adaptive subspace codebook outperforms the conventional channel statistics-based codebook in terms of the per-user rate.
This is caused by the fact that the proposed AoD-adaptive subspace codebook concentrates the quantization vectors exactly on the channel subspace.

\begin{figure}[tb!]
	\vspace*{-1mm}
	\begin{center}
		\includegraphics[width=0.7\columnwidth, keepaspectratio]{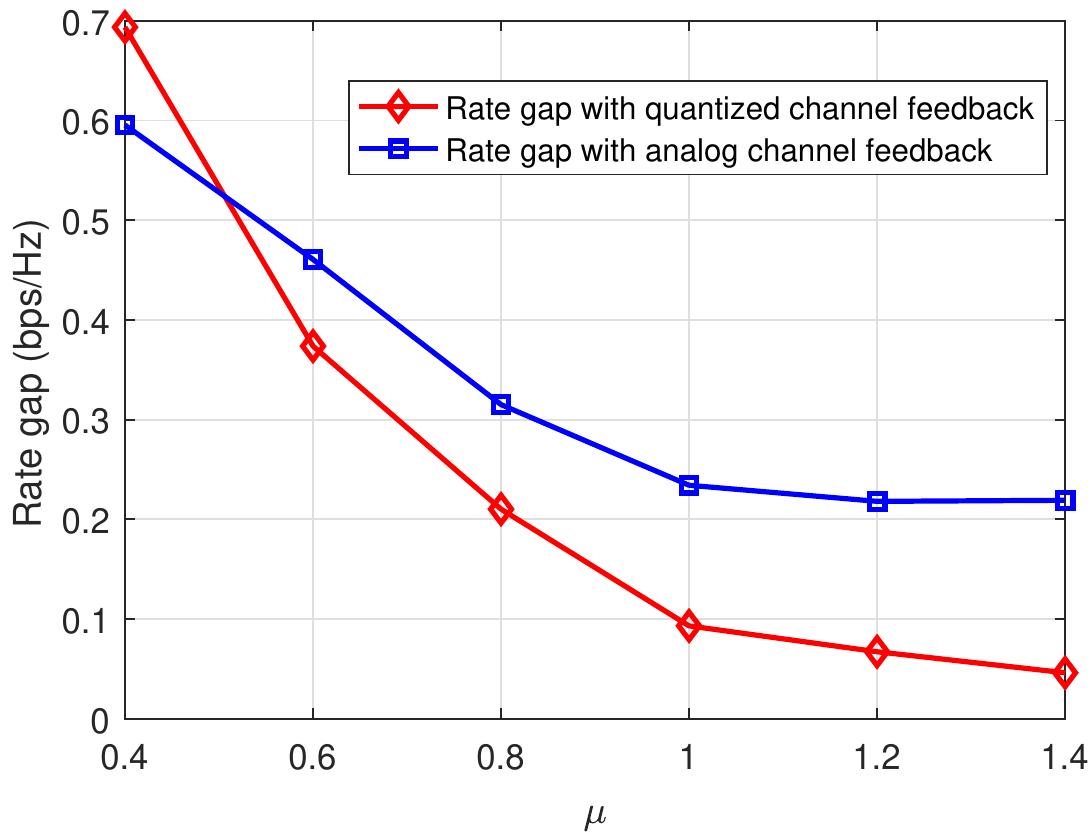}
	\end{center}
	\vspace*{-4mm}
	\caption{The rate gap $\Delta R_\text{Quanized}$ with quantized channel feedback using the proposed AoD-adaptive subspace codebook and the rate gap $\Delta R_\text{Analog}$ with analog channel feedback against $\mu$ with a constant uplink SNR $\gamma_\text{U}=5$. Quantized channel feedback with the proposed codebook outperforms analog channel feedback when the uplink channel rate is larger than the source rate, i.e., $\mu$ is large.}
	\label{Fig8}
	\vspace*{-1mm}
\end{figure}

Finally, we compare quantized channel feedback using the proposed AoD-adaptive subspace codebook with analog channel feedback in terms of rate gap. Fig. \ref{Fig8} shows the rate gap comparison between quantized channel feedback and analog channel feedback against the scale factor $\mu$ when a constant uplink SNR $\gamma_\text{U}=5$ is configured. 
The downlink receiver is 10 dB.
We observe that the rate gap with quantized channel feedback decreases (exponentially) more quickly than that with analog channel feedback (inversely). That is to say, quantized channel feedback outperforms analog channel feedback when $\mu$ is large, which is consistent with the \textbf{Remark 2} in Section \ref{S5}. 

\begin{figure}[tb!]
	\vspace*{-1mm}
	\begin{center}
		\includegraphics[width=0.7\columnwidth, keepaspectratio]{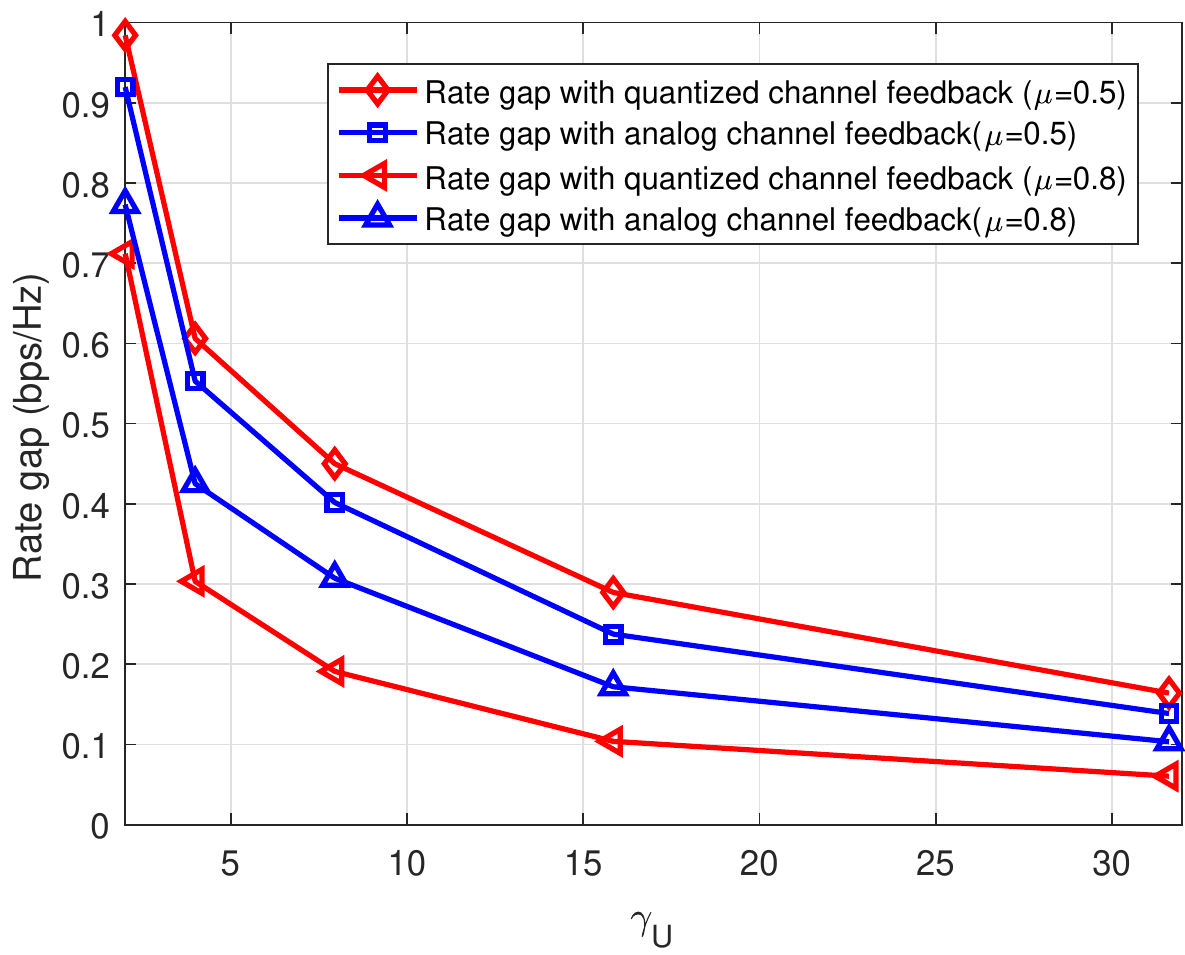}
	\end{center}
	\vspace*{-4mm}
	\caption{The rate gap $\Delta R_\text{Quanized}$ with quantized channel feedback using the proposed AoD-adaptive subspace codebook and the rate gap $\Delta R_\text{Analog}$ with analog channel feedback against uplink SNR $\gamma_\text{U}$ when $\mu =0.5 \le \frac{P-1}{P}$ and $\mu = 0.8 \ge \frac{P-1}{P}$ are configured, respectively.}
	\label{Fig9}
	\vspace*{-1mm}
\end{figure}

Fig. \ref{Fig9} shown the rate gap comparison between quantized channel feedback and analog channel feedback against the uplink SNR $\gamma_\text{U}$ when $\mu =0.5 \le \frac{P-1}{\textsc{P}}$ and $\mu = 0.8 \ge \frac{P-1}{P}$ are configured, respectively. The downlink receiver is 10 dB. We observe that the rate gap of quantized channel feedback is larger than that of analog channel feedback when $\mu =0.5 \le \frac{P-1}{P}$ is configured. 
While the rate gap of quantized channel feedback is smaller than that of analog channel feedback when $\mu =0.8 \ge \frac{P-1}{P}$ is configured. These simulated results are consistent with the \textbf{Remark 3} in section \ref{S5}.

\section{Conclusions}\label{S6}
In this paper, we have proposed the AoD-adaptive subspace codebook for channel feedback in FDD massive MIMO systems.
By exploiting the property that path AoDs vary more slowly than the path gains, the proposed codebook can achieve significant reduction of codebook size and feedback overhead.
We have also provided performance analysis of the proposed codebook in the large-dimensional regime,
where we have proved that the required number of feedback bits only scales linearly with the number of paths,
which is much smaller than the number of BS antennas.
Moreover, we have compared quantized channel feedback using the proposed AoD-adaptive codebook with analog channel feedback.
These analytical results are verified by extensive simulations.

\section*{Appendix I}\label{S7.1}
In this Apppendix, we prove the following \textbf{Lemma 3} in the large-dimensional regime which is used in the proof of \textbf{Lemma 2} as well as in the analysis of quantization error in Section \ref{S4.2}.

\textbf{Lemma 3}:
The steering vectors of paths with different AoDs (column vectors of $\mathbf{A}$) are asymptotically orthogonal to each other when $M\rightarrow\infty$, i.e.,
$\mathbf{A}^\text{H}\mathbf{A} \overset{M\rightarrow\infty}{=} \mathbf{I}_P$.
When the quantization error of the AoD is small, $\mathbf{A}^\text{H}\hat{\mathbf{A}} \overset{M\rightarrow\infty}{=} \mathcal{K}\mathbf{I}_P$
where $\left|\mathcal{K}\right|^2 \geq1-\frac{M^2}{3}(\pi\frac{d}{\lambda})^2r^22^{-2B_0}$.
\begin{IEEEproof}
For the ULA, the $(p,q)$-th element of $\mathbf{A}^\text{H}\hat{\mathbf{A}}$ is $\mathbf{a}(\phi_p)^\text{H}\mathbf{a}(\hat{\phi_q})= \frac{1}{M}\sum_{m=0}^{M-1}e^{-j2\pi\frac{d}{\lambda}m[\sin(\phi_p)-\sin(\hat{\phi_q})]}$.
Denoting $\delta_{p,q}=\sin(\phi_p)-\sin(\hat{\phi_q})$, we have
\begin{align} \label{eq_apaqH}
\left|\mathbf{a}(\phi_p)^\text{H}\mathbf{a}(\hat{\phi_q})\right|=\left|\Upsilon(\frac{d}{\lambda}\delta_{p,q})\right|,
\end{align}
where $\Upsilon(x)\triangleq\frac{\sin(M\pi x)}{M\sin(\pi x)}$.
According to the characteristics of $\Upsilon(x)$,
when $|x|\gg\frac{1}{M}$, i.e., $|\delta_{p,q}|\gg\frac{\lambda}{Md}$, we have $|\Upsilon(x)|\overset{M\rightarrow\infty}{=} 0$ \cite{CL_XGao_BeamSelection}.
Now we consider the following two cases:

i) If $\hat{\mathbf{A}}=\mathbf{A}$, i.e., $\delta_{p,q}=\sin\phi_p-\sin\phi_q$,
the absolute diagonal element $|\mathbf{a}(\phi_p)^\text{H}\mathbf{a}(\phi_p)|=|\Upsilon(\frac{d}{\lambda}\delta_{p,p})|=|\Upsilon(0)|=1$.
For the non-diagonal element $\mathbf{a}(\phi_p)^\text{H}\mathbf{a}(\phi_q)$,
since AoDs $\phi_p$ and $\phi_q$ are distinguished enough,
i.e., $|\delta_{p,q}|=|\sin\phi_p-\sin\phi_q|\gg\frac{\lambda}{Md}$,
the absolute non-diagonal element $|\mathbf{a}(\phi_p)^\text{H}\mathbf{a}(\phi_q)|=|\Upsilon(\frac{d}{\lambda}\delta_{p,q})| \overset{M\rightarrow\infty}{=} 0$.
Therefore, we have $\mathbf{A}^\text{H}\mathbf{A} \overset{M\rightarrow\infty}{=} \mathbf{I}_P$.

ii) Otherwise $\hat{\mathbf{A}}\neq\mathbf{A}$, i.e., $\delta_{p,q}=\sin\phi_p-\sin\hat{\phi}_q$,
the absolute diagonal element $|\mathbf{a}(\phi_p)\mathbf{a}(\hat{\phi}_p)^\text{H}|=|\Upsilon(\frac{d}{\lambda}\delta_{p,p})|$,
where $|\delta_{p,p}|=|\sin\phi_p-\sin\hat{\phi_p}|$ is the AoD quantization error.
With uniform quantization, $|\delta_{p,p}|\leq r2^{-B_0}$,
where $r$ is the difference between the maximum and minimum values over which $\sin(\phi_p)$ is quantized and $B_0$ is the number of quantization bits.
By denoting $\mathcal{K}=\mathbf{a}(\phi_p)^\text{H}\mathbf{a}(\hat{\phi_p})$ and using (\ref{eq_apaqH}), we have
\begin{align}\label{eq_U_M}
\left|\mathcal{K}\right|^2&=\frac{\sin^2(\pi\frac{d}{\lambda}\delta_{p,p}M)}{M^2\sin^2(\pi\frac{d}{\lambda}\delta_{p,p})}\\\nonumber
&\overset{(a)}{\approx} 1-\frac{M^2}{3}\left(\pi\frac{d}{\lambda}\right)^2\delta_{p,p}^2\\\nonumber
&\overset{(b)}{\geq} 1-\frac{M^2}{3}\left(\pi\frac{d}{\lambda}\right)^2r^22^{-2B_0},
\end{align}
where (a) is obtained by the second order Taylor's expansion of $\sin^2(\pi\frac{d}{\lambda}\delta_{p,p}M)=\left( \pi\frac{d}{\lambda}\delta_{p,p}M\right) ^2-\left( \pi\frac{d}{\lambda}\delta_{p,p}M\right)^4/3 $ and  the first order Taylor's expansion of $\sin^2(\pi\frac{d}{\lambda}\delta_{p,p})=\left( \pi\frac{d}{\lambda}\delta_{p,p}\right) ^2$.
(b) holds true as $|\delta_{p,p}|\leq r2^{-B_0}$.
For the non-diagonal element $\mathbf{a}(\phi_p)^\text{H}\mathbf{a}(\hat{\phi}_q)$,
\begin{align}
|\delta_{p,q}|&=|\sin\phi_p-\sin\hat{\phi_q}| \\\nonumber
&=|\sin\phi_p-\sin\phi_q+\sin\phi_q-\sin\hat{\phi_q}|\\\nonumber
&\geq |\sin\phi_p-\sin\phi_q| - |\sin\phi_q-\sin\hat{\phi_q}|\\\nonumber
&\overset{(a)}{\geq}|\sin\phi_p-\sin\phi_q|-r2^{-B_0},
\end{align}
where (a) is true since $|\sin\phi_q-\sin\hat{\phi_q}|\leq r2^{-B_0}$.
We assume that $B_0$ is properly chosen (large enough).
Then, we can obtain $|\delta_{p,q}|\gg \frac{\lambda}{dM}$.
Thus, it holds that the absolute non-diagonal element $|\mathbf{a}(\phi_p)^\text{H}\mathbf{a}(\hat{\phi}_q)|=|\Upsilon(\frac{d}{\lambda}\delta_{p,q})|\overset{M\rightarrow\infty}{=} 0$.
Therefore, we have $\mathbf{A}^\text{H}\hat{\mathbf{A}}\overset{M\rightarrow\infty}{=} \mathcal{K}\mathbf{I}_P$.

In summary, $\mathbf{A}^\text{H}\mathbf{A}\overset{M\rightarrow\infty}{=} \mathbf{I}_P$, and $\mathbf{A}^\text{H}\hat{\mathbf{A}}\overset{M\rightarrow\infty}{=} \mathcal{K}\mathbf{I}_P$
where $\left|\mathcal{K}\right|^2 \geq1-\frac{M^2}{3}(\pi\frac{d}{\lambda})^2r^22^{-2B_0}$.

For the UPA, by denoting $\mathbf{a}_h(\phi,\theta)=\dfrac{1}{\sqrt{M_1}}\left[1,e^{j2\pi\frac{d}{\lambda}\cos\theta\sin\phi},\cdots, e^{j2\pi\frac{d}{\lambda}(M_1-1)\cos\theta\sin\phi}\right]^\text{T}$ and \\ $\mathbf{a}_v(\phi,\theta)=\dfrac{1}{\sqrt{M_2}}\left[1,e^{j2\pi\frac{d}{\lambda}\sin\theta},\cdots, e^{j2\pi\frac{d}{\lambda}(M_2-1)\sin\theta}\right]^\text{T}$, we can express the antenna array response for the UPA in (\ref{eq_ste_vec_UPA}) as $\mathbf{a}(\phi,\theta)=\mathbf{a}_h(\phi,\theta)\otimes\mathbf{a}_v(\phi,\theta)$. 
Similar to the case of ULA, we calculate
\begin{align} \mathbf{a}(\phi_p,\theta_p)^\text{H}\mathbf{a}(\hat{\phi}_q,\hat{\theta}_q)= \left( \mathbf{a}_h(\phi_p,\theta_p)^\text{H}\mathbf{a}_h(\hat{\phi}_q,\hat{\theta}_q)\right)  \otimes \left( \mathbf{a}_v(\phi_p,\theta_p)^\text{H}\mathbf{a}_v(\hat{\phi}_q,\hat{\theta}_q)\right). 
\end{align}
By denoting $\zeta_{p,q}=\cos\theta_p\sin\phi_p-\cos\hat{\theta_q}\sin\hat{\phi}_q$ and $\xi_{p,q} =\sin\theta_p - \sin\hat{\theta}_q$, we have
\begin{align}
\left|\mathbf{a}(\phi_p,\theta_p)^\text{H}\mathbf{a}(\hat{\phi}_q,\hat{\theta}_q)\right|=\left|\Upsilon(\frac{d}{\lambda}\zeta_{p,q})\right|\left|\Upsilon(\frac{d}{\lambda}\xi_{p,q})\right|.
\end{align}
Now we consider the following two cases:

i) If $\hat{\mathbf{A}}=\mathbf{A}$, i.e., $\zeta_{p,q} = \cos\theta_p\sin\phi_p-\cos\theta_q\sin\phi_q$ and $\xi_{p,q}=\sin\theta_p-\sin\theta_q$, the absolute diagonal element $\left|\mathbf{a}(\phi_p,\theta_p)^\text{H}\mathbf{a}(\phi_p,\theta_p)\right|=|\Upsilon(\frac{d}{\lambda}\zeta_{p,p})\|\Upsilon(\frac{d}{\lambda}\xi_{p,p})|=|\Upsilon(0)\|\Upsilon(0)|=1$. 
For the non-diagonal element $|\mathbf{a}(\phi_p,\theta_p)^\text{H}\mathbf{a}(\phi_q,\theta_q)|=|\Upsilon(\frac{d}{\lambda}\zeta_{p,q})\|\Upsilon(\frac{d}{\lambda}\xi_{p,q})|\overset{M\rightarrow\infty}{=} 0$, since the azimuth (elevation) AoDs $\phi_p$ ($\theta_p$) and $\phi_q$($\theta_q$) are distinguished enough. 
Therefore, we have $\mathbf{A}^\text{H}\mathbf{A} \overset{M\rightarrow\infty}{=} \mathbf{I}_P$.

ii) Otherwise $\hat{\mathbf{A}}\neq\mathbf{A}$, i.e., $\zeta_{p,q} = \cos\theta_p\sin\phi_p-\cos\hat{\theta}_q\sin\hat{\phi}_q$ and $\xi_{p,q}=\sin\theta_p-\sin\hat{\theta}_q$. For the diagonal element,
we have 
\begin{align}
\left|\mathcal{K}\right|^2&= |\Upsilon(\frac{d}{\lambda}\zeta_{p,p})|^2|\Upsilon(\frac{d}{\lambda}\xi_{p,p})|^2\\\nonumber
&=\frac{\sin^2(\pi\frac{d}{\lambda}\zeta_{p,p}M_1)}{M_1^2\sin^2(\pi\frac{d}{\lambda}\zeta_{p,p})}\times\frac{\sin^2(\pi\frac{d}{\lambda}\xi_{p,p}M_2)}{M_2^2\sin^2(\pi\frac{d}{\lambda}\xi_{p,p})}\\\nonumber
&{\approx} \left( 1-\frac{M_1^2}{3}\left(\pi\frac{d}{\lambda}\right)^2\zeta_{p,p}^2\right)\left( 1-\frac{M_2^2}{3}\left(\pi\frac{d}{\lambda}\right)^2\xi_{p,p}^2\right).
\end{align}
For the non-diagonal element, similar to the ULA, when the number of bits for AoD quantization is properly chosen, we have $|\mathbf{a}(\phi_p,\theta_p)^\text{H}\mathbf{a}(\hat{\phi}_q,\hat{\theta}_q)|=|\Upsilon(\frac{d}{\lambda}\zeta_{p,q})\|\Upsilon(\frac{d}{\lambda}\delta_{p,q})|\overset{M\rightarrow\infty}{=} 0$. 
Therefore, we have $\mathbf{A}^\text{H}\hat{\mathbf{A}}\overset{M\rightarrow\infty}{=} \mathcal{K}\mathbf{I}_P$.

In summary, $\mathbf{A}^\text{H}\mathbf{A}\overset{M\rightarrow\infty}{=} \mathbf{I}_P$, and $\mathbf{A}^\text{H}\hat{\mathbf{A}}\overset{M\rightarrow\infty}{=} \mathcal{K}\mathbf{I}_P$
where $\left|\mathcal{K}\right|^2 \approx \left( 1-\frac{M_1^2}{3}\left(\pi\frac{d}{\lambda}\right)^2\zeta_{p,p}^2\right)\\\left( 1-\frac{M_2^2}{3}\left(\pi\frac{d}{\lambda}\right)^2\delta_{p,p}^2\right)$.
\end{IEEEproof}

\section*{Appendix II}\label{S7.2}
In this Appendix, we prove the following \textbf{Lemma 4} which is used in the proof of \textbf{Lemma 2}.

\textbf{Lemma 4}:
In the extreme case that all $U$ users share the same clusters around the BS, i.e., $P_1=P_2=,\cdots,=P_U=P$ and $\mathbf{A}_1 =\mathbf{A}_2=,\cdots, =\mathbf{A}_U=\mathbf{A}$, the subscript $u$ of $P_u$, $\mathbf{A}_u$ and $\hat{\mathbf{A}}_u$ can be omitted.
We have $\text{E}\left[|\mathbf{t}^\text{H}\mathbf{u}|^2\right] =\frac{1}{P-1}$.
\begin{IEEEproof}
Based on Section \ref{S3}, we have $\mathbf{q}=\mathbf{A}\mathbf{t}$ and
the fed back channel vector can be expressed as $\hat{\mathbf{h}}_u=\|\mathbf{h}_u\|\hat{\mathbf{A}}\mathbf{w}_{i_u}$.
Considering $\mathbf{q}$ is distributed in the null space of $\hat{\mathbf{h}}_u$ in \textbf{Lemma 2}, we get
\begin{align}
\mathbf{q}^\text{H}\hat{\mathbf{h}}_u&=\mathbf{t}^\text{H}\mathbf{A}^\text{H}\|\mathbf{h}_u\|\hat{\mathbf{A}}\mathbf{w}_{i_u}=0,
\end{align}
where $\mathbf{A}^\text{H}\hat{\mathbf{A}}\overset{M\rightarrow\infty}{=} \mathcal{K}\mathbf{I}_{P}$ according to Appendix I.
Therefore, we have $\mathbf{t}^\text{H}\mathbf{w}_{i_u}=0$,
i.e., $\mathbf{t}$ is isotropically distributed in the null space of $\mathbf{w}_{i_u}$.

On the other hand, based on Section \ref{S3},
the precoding vector can be expressed as $\mathbf{\hat{v}}_i=\mathbf{A}\mathbf{u}$,
and the fed back channel vector can be expressed as $\hat{\mathbf{h}}_u=\|\mathbf{h}_u\|\hat{\mathbf{A}}\mathbf{w}_{i_u}$.
Since we use the ZF precoding,
$\mathbf{\hat{v}}_i$ is orthogonal to $\hat{\mathbf{h}}_u$, i.e.,
\begin{align}
\mathbf{\hat{v}}_i^\text{H}\hat{\mathbf{h}}_u&=\mathbf{u}^\text{H}\mathbf{A}^\text{H}\|\mathbf{h}_u\|\hat{\mathbf{A}}\mathbf{w}_{i_u}=0,
\end{align}
where $\mathbf{A}^\text{H}\hat{\mathbf{A}}\overset{M\rightarrow\infty}{=} \mathcal{K}\mathbf{I}_{P}$ according to Appendix I.
Therefore, we have $\mathbf{u}^\text{H}\mathbf{w}_{i_u}=0$,
i.e., $\mathbf{u}$ is isotropically distributed in the null space of $\mathbf{w}_{i_u}$.

Now we have proved that both $\mathbf{t}$ and $\mathbf{u}$ are unit-norm isotropic vectors in the null space of $\mathbf{w}_{i_u}$.
Based on \cite{TIT_NJindal_MIMOBroadcast}, we have
\begin{align}
\text{E}\left[|\mathbf{t}^\text{H}\mathbf{u}|^2\right] =\frac{1}{P-1}.
\end{align}
\end{IEEEproof}
\bibliographystyle{IEEEtran}
\bibliography{IEEEabrv,Gao1Ref}
\end{document}